\pdfoutput=1
\documentclass[aps,prd,amsmath,floats,floatfix,superscriptaddress,nofootinbib,showpacs]{revtex4-2}

\usepackage[T1]{fontenc}
\usepackage[utf8]{inputenc}
\usepackage{lmodern}
\usepackage{amsmath}
\usepackage{lipsum}

\usepackage[dvipsnames, usenames]{xcolor}
\definecolor{linkcolor}{rgb}{0.0,0.3,0.5}
\usepackage[hypertexnames=false, unicode, colorlinks=true, linkcolor=linkcolor,
citecolor=linkcolor, filecolor=linkcolor,urlcolor=linkcolor,
pdfusetitle]{hyperref}

\usepackage[all]{hypcap}
\usepackage{graphicx}
\usepackage{xspace}

\usepackage{amssymb}
\usepackage[normalem]{ulem} 
\usepackage{bm} 

\usepackage{microtype}

\usepackage[english]{babel}
\usepackage{blindtext}

\renewcommand{\vec}[1] {\bm{#1}}

\newcommand*{\la}  {\left\langle}
\newcommand*{\ra}  {\right\rangle}

\definecolor{darkteal}{RGB}{0,111,111}

\begin{document}

\rightline{\scriptsize RBI-ThPhys-2025-09}

\title{Projecting Unequal Time Fields and Correlators of Large Scale Structure}

\author{Theodore Steele}
\affiliation{Astronomy Centre, University of Sussex, Brighton, BN1 9RH, UK}

\author{Alvise Raccanelli}
\affiliation{Dipartimento di Fisica Galileo Galilei, Università di Padova, I-35131 Padova, Italy}
\affiliation{INFN Sezione di Padova, I-35131 Padova, Italy}
\affiliation{INAF-Osservatorio Astronomico di Padova, Italy}

\author{Zvonimir Vlah}
\affiliation{Ruđer Bošković Institute, Bijenička cesta 54, 10000 Zagreb, Croatia}
\affiliation{Kavli Institute for Cosmology, University of Cambridge, Cambridge CB3 0HA, UK}
\affiliation{Department of Applied Mathematics and Theoretical Physics, University of Cambridge, Cambridge CB3 0WA, UK}

\begin{abstract}
Many large scale structure surveys sort their observations into redshift bins and treat every tracer as being located at the mean redshift of its bin, a treatment which we refer to as the equal time approximation.  Recently, a new method was developed which allows for the estimation and correction of errors introduced by this approximation, which we refer to as the unequal time correlator-level projection. For single tracer power spectra, corrections arise at second order and above in a series expansion, with first order terms surviving only in multi-tracer analyses.  In this paper we develop a new method which we refer to as the unequal time field level projection. This formalism projects the fields individually onto the celestial sphere, displaced from individual reference times, before defining their correlators.  This method introduces new, first order correction terms even in the case of single tracer power spectra. Specifically, new first order terms are introduced which apply to both cross-bin and single bin correlators.  All of these new corrections originate with derivatives over combinations of a delta function, a cross-bin phase term, and the power spectrum itself and stem from the introduction of two unequal time Fourier transforms into the analysis.  We analyse these corrections in the context of a linearly biased power spectrum divided between two redshift bins and find that they can lead to non-trivial corrections, particularly to cross-bin correlators.  We also show that these terms can be replicated by appropriately extending the correlator-level analysis to include a second Fourier transform which allows for a full redshift bin integration.  
\end{abstract}

\maketitle

\section{Introduction}
\label{sec:intro}
Current and near future surveys of cosmological Large Scale Structure (LSS) such as EUCLID \cite{2011arXiv1110.3193L}, SPHEREx \cite{Dore:2014cca}, the Vera C. Rubin Observatory surveys \cite{2019ApJ...873..111I}, HSC \cite{Hamana:2019etx}, SKA \cite{10.1093/mnras/stv695}, DESI \cite{Wang:2021moa}, WFIRST \cite{interloperbias}, 4MOST \cite{4most}, PFS \cite{10.1093/pasj/pst019}, KiDS \cite{Hildebrandt:2020rno}, BOSS \cite{2013AJ....145...10D}, and eBOSS \cite{10.1093/mnras/stz3602} will provide enough data to allow for an unprecedented level of precision in the cosmological analysis of fundamental physics, such as the properties of dark energy \cite{2013PhR...530...87W,Euclid:2021cfn,Piga:2022mge,2011PhRvD..84h3501H}, primordial non-Gaussianity \cite{2012MNRAS.422.2854G,Karagiannis:2018jdt,Yamauchi:2014ioa,Brown:2024dmv}, neutrino masses \cite{Hahn:2019zob,Cerbolini:2013uya,Euclid:2022qde,Dvorkin:2019jgs}, and the assumption of approximate homogeneity and isotropy on large scales \cite{Euclid:2021frk,2023A&A...671A..68C,Pandey:2021qqp,Zunckel:2010yq}.

LSS surveys study the overall matter distribution of the Universe by observing tracers of that distribution, such as the number density field of galaxies or the integrated effect of weak gravitational lensing.  Primarily, the data is analysed using the density contrast field, the normalised difference between the observed density at a given point in space and time and the mean density at that time.

In order to use measured data to place constraints on fundamental physics, observations from LSS surveys are studied through statistical objects such as correlators, the expectation values of products of density contrast fields.  By theoretically modeling these objects \cite{Bernardeau:2001qr,Blas:2015qsi,Blas:2016sfa,Blas:2015tla,Baumann:2010tm,Carrasco:2012cv,Arico:2021izc} as functions of the parameters describing fundamental physics \cite{Cusin:2017wjg,Assassi:2015jqa,Senatore:2017hyk,Chudaykin:2019ock,Vasudevan:2019ewf} and comparing these models to survey measurements \cite{Zhang:2021uyp,Ivanov:2019pdj,Carrilho:2022mon,DESI:2024mwx,Chen:2024vuf,Cabass:2024wob,Sugiyama:2023tes}, constraints can be placed on those parameters.  Before such calibrations can be made, however, we must note that theoretical models of LSS are generally studied in what we call Hyperuranion space, which is a time evolving 3D grid in which the exact location of any given object at any given time can be known with certainty.  Of course, surveys are limited to being able to observe the Universe in a spherical grid centered on the telescope and, due to the finite speed of light, cannot separate their radial coordinate from time.  The projection from Hyperuranion space to this spherical system, which we refer to as the celestial sphere, and the effects which manifest in this operation but were hitherto undiscovered constitute the subject of this paper.

When analyzing their data, LSS surveys often divide their radial coordinate into redshift bins \cite{DES:2017ndt,Euclid:2021upd,2012AJ....144..144B} and treat every tracer located within a given bin as though it were located at the mean redshift of that bin, an assumption which we refer to as the equal time approximation.  This is convenient and it is usually assumed that any errors introduced by this approximation will be below the precision limitation of the survey.  However, as we enter an era of unprecedentedly precise cosmology, it is becoming important to test and potentially correct such assumptions.  

There are multiple possible approaches to the problem of unequal time effects in cosmology, including formalisms for introducing corrections to otherwise equal time power spectra \cite{Scoccimarro:1997st} as well as spectra which have been explicitly integrated over the light cone \cite{Pryer:2021cut,Semenzato:2024rlc}.  In a recent series of papers~\cite{Raccanelli:2023fle,Raccanelli:2023zkj,Gao:2023rmo,Gao:2023tcd}, a formalism was derived to introduce corrections to otherwise equal time power spectra which we refer to as the unequal time correlator-level projection.  In this formalism, correlators of tracers located at different times were defined in Hyperuranion space via a Taylor expansion around their mean time.  A new projection operation was then defined which allowed these unequal time correlators to be projected onto the celestial sphere in a manner that preserved information regarding their density fields' unequalness in time.  It was found that multi-tracer power spectra exhibit first order correction terms originating from their Taylor expansion, while single tracer power spectra only exhibit corrections at second order and above.

In this paper, we define what we call the field level projection.  In this formalism, the time of observation of each density field is defined by a Taylor expansion around the mean redshift of the bin into which it is sorted.  We then project each field individually in a manner that preserves information relating to this time displacement and only correlate them once they have already been projected onto the celestial sphere.  The field level projection formalism allows for the study of correlators within a given bin or split between bins with equal efficiency.  We find that three new correction terms are introduced at first order in the context of a single tracer power spectrum; one of these accounts for the evolution of the matter field between different bins, one for the evolution of the biasing between those same bins, and one for the displacement of the fields from their bins' mean redshifts.  We show that these terms can lead to non-trivial corrections which were hitherto unaccounted for, both within single bin spectra and in the case of cross-bin analyses.  These terms arise due to the presence of two Fourier analogues of unequalness in time parameters, allowing the two fields to be individually displaced from the mean redshift of their bin.  In the correlator-level projection that has been studied before, only a single unequalness in time Fourier transform is defined and this leads to the two field's having their redshift displacements cancel one another out due to being equally displaced on opposite sides of the reference redshift.  In an appendix, we show that the single bin corrections can be rederived in the context of an appropriately extended version of the correlator-level projection which incorporates a second Fourier transform to integrate our correlators themselves to be displaced from the mean redshift of the bin.

This paper is structured as follows: in Sec.~\ref{sec:StatObs}, we summarise known results on the definitions of correlators and their role in modern cosmology.  In Sec.~\ref{sec:TimeExpansion}, we rederive a formalism for describing time displaced density fields in Hyperuranion space first developed in \cite{Raccanelli:2023fle}.  In Sec.~\ref{sec:Projecting} we review the well known notion of projecting a power spectrum onto the celestial sphere and discuss the correlator-level projection.  In Sec.~\ref{sec:FLP} we introduce the field level projection and analyse the newly discovered first order corrections in the context of a linearly biased, single tracer power spectrum split between two redshift bins. In Sec.~\ref{sec:discussion}, we review our results and discuss their implications for cosmology.  In App.~\ref{sec:BACL}, we develop an extension of the correlator-level projection formalism and show that it replicates the results of a single bin, field level projection.  We note that throughout this paper we are working in natural units such that $c=1$.

\section{Cosmological Statistics}
\label{sec:StatObs}
The study of both the CMB and LSS is statistical in nature.  In order to place constraints on physical parameters, we describe the statistical distribution of matter and its tracers as a function of those parameters and fit their values to observations.  The primary statistical objects used in such studies are correlation functions of density contrast fields and their Fourier space counterparts, spectra, both of which we collectively refer to as correlators.  We define $\mathcal{S}_{n}$, the spectrum of $n$ density contrast fields, as
\begin{equation}
\la\prod_{m=1}^{n}\delta(\mathbf{k}_{m},\chi_{m})\ra\equiv (2\pi)^{3}\delta^{\mathrm{3D}}\left(\sum_{m=1}^{n}\mathbf{k}_{m}\right)\mathcal{S}_{n}(\mathbf{k}_{1},\ldots,\mathbf{k}_{n},\chi_{1},\ldots,\chi_{n})~,
    \label{eq:CorrSpec}
\end{equation}
where the density fields can represent different combinations of tracers, the delta function accounts for cosmological symmetries, and we note that the comoving distance is defined as
\begin{equation}
    \chi(z)\equiv\int_{0}^{z}\frac{dz'}{H(z')}~.
\end{equation}
In the case where we are correlating two fields, we refer to our spectrum as the power spectrum and label it $\mathcal{S}_{2}(k,\chi_{1},\chi_{2})=\mathcal{P}(k,\chi_{1},\chi_{2})$, where we may note the implementation of the delta function from Eq.~\eqref{eq:CorrSpec} and the scalar dependency upon $k$ due to assumed symmetries.

The definitions above can be applied to the underlying gravitating matter structure of the Universe.  However, directly observable tracers such as galaxies follow a slightly modified statistical distribution.  Due to the mutual attraction between light and dark matter, galaxy distributions are strongly correlated with underlying dark matter distributions, but their different masses and interactions with other forces lead to the requirement for corrections.  The differences between observable tracer distributions and underlying matter distributions are referred to as biasing \cite{Desjacques:2016bnm} and may be accounted for by including appropriate parameters into the definitions of our density fields: $\delta(\mathbf{k},\chi)\rightarrow b(\chi)\delta(\mathbf{k},\chi)$, where we note the redshift dependence of biasing parameters which allow for the evolution of the disparity between light and dark matter distributions with time.

The most commonly studied correlators are spectra of a single tracer, particularly of matter fields through the study of lensing and galaxy fields through direct observations \cite{10.1093/mnras/285.1.L5,2010A&A...523A..28K,2024A&A...684A.138E}.  However, cross-correlators \cite{10.1111/j.1365-2966.2008.13371.x,Fraser:2024ecp,Kirk:2015xqa} and correlators of other observables such as matter velocity fields \cite{Dam:2021fff,Howlett:2017asq,Tonegawa:2023gbf} are becoming increasingly important as they often contain additional information and can be used to break degeneracies that arise in matter and galaxy autocorrelator analyses \cite{Kacprzak:2022oit,Singh:2018kmr,Schmittfull:2017ffw}.

\section{Time Expansion}
\label{sec:TimeExpansion}
In this section, we rederive an expression for time displacement in Hyperuranion linear density fields that was first explored in~\cite{Raccanelli:2023fle}.  We do this, rather than accounting for differences in time by directly giving different values to the $\chi_{m}$ in Eq.~\eqref{eq:CorrSpec}, in order to obtain a formalism which will permit the averaging of small displacements from assumed redshifts and the preservation of this displacement when projected onto the celestial sphere.  

We wish to describe a density contrast field which is at some comoving distance $\chi$ that is displaced by $\delta\chi$ from the position $\bar{\chi}$:
\begin{equation}
\chi\equiv\bar{\chi}+\delta\chi~.
\label{eq:chi}
\end{equation}
$\bar{\chi}$ could be the mean of the redshift bin into which an observable at comoving distance $\chi$ has been sorted, the mean comoving distance of two radially separated observables being correlated, or any other chosen reference distance. For the rest of this paper, we will work with two formalisms, one in which it is taken to be the mean redshift of a given correlator, which we make use of in the correlator-level analysis, and one in which it is taken to be the mean redshift of a redshift bin, which we make use of in the field level analysis.

To derive a series expansion for a linear density field located at $\chi$, we begin by Taylor expanding around $\bar{\chi}$:
\begin{equation}
\delta(\mathbf{k},\chi)=\delta(\mathbf{k},\bar{\chi})+\frac{d}{d\chi}\delta(\mathbf{k},\chi)\bigg|_{\delta \chi=0}\delta \chi
+\frac{1}{2}\frac{d^2}{d\chi^{2}}\delta(\mathbf{k},\chi)\bigg|_{\delta\chi=0}\delta\chi^2+\dots.
\label{eq:TExpansion}
\end{equation}
We can now reparametrise our derivatives:
\begin{equation}
\frac{d}{d\chi}\delta(\mathbf{k},\chi) = \frac{dz}{d\chi}\frac{d}{dz}\delta(\mathbf{k},\chi) = F_{a}(\chi)H(\chi)\delta(\mathbf{k},\chi)~,
\label{eq:Taylorstep}
\end{equation}
where we have used the fact that for a linear density field, $
\partial_{z}\delta(\mathbf{k},\chi)=-f(\chi)a(\chi)\delta(\mathbf{k},\chi) \,$
for the logarithmic growth rate $f(\chi)$ and scale factor $a(\chi)$ and have defined
\begin{equation}
    F_{a}(\chi)\equiv -f(\chi)a(\chi)~.
    \end{equation}
Thus, up to first order in the time expansion, our field is given by
\begin{equation}
\delta(\mathbf{k},\chi)=\delta(\mathbf{k},\bar{\chi})\left[1+F_{a}(\bar{\chi})H(\bar{\chi})\delta\chi_{m}\right]~.
\label{eq:expandeddelta}
\end{equation}

If we are considering biased tracers by introducing biasing parameters and repeating the above analysis, Eq.~\eqref{eq:Taylorstep} becomes
\begin{equation}
\label{eq:biasexp}
\frac{d}{d\chi}b(\chi)\delta(\mathbf{k},\chi) = \frac{dz}{d\chi}\frac{d}{dz}b(\chi)\delta(\mathbf{k},\chi) = F_{a}(\chi)H(\chi)\delta(\mathbf{k},\chi)+\partial_{z}b(\chi)H(\chi)\delta(\mathbf{k},\chi)
\end{equation}
and we find that our biased analogue of Eq.~\eqref{eq:expandeddelta} is
\begin{equation}
b(\chi)\delta(\mathbf{k},\chi)=\delta(\mathbf{k},\bar{\chi})\left[b(\bar{\chi})+\left(F_{a}(\chi)b(\bar{\chi})+\partial_{z}b(\bar{\chi})\right)H(\bar{\chi})\delta\chi_{m}\right]~.
\label{eq:biasedexpansion}
\end{equation}
We note that this differs from Eq.~\eqref{eq:expandeddelta} not only in the multiplication of the existing terms by the biasing parameter but also by the inclusion of the derivative of that parameter.  

In this section, we have shown how we can derive expressions for linear density fields displaced by a small distance from a defined redshift in a manner that provides explicit multiplicative corrections to correlators already estimated at those defined redshifts.  These corrections account for the evolution of both the dark matter distribution and its relation to the overlaying galaxy number distribution.  Since the Universe is constantly evolving, the assumption that all such fields can be treated as being located at a single time has been shown to lead to errors proportional to the rate of change with redshift of the parameters representing both dark matter and biasing.  This analysis has taken place in Hyperuranion space and has not accounted for projection effects that will become relevant in the remainder of this paper.   

\section{Image Projection}
\label{sec:Projecting}
In this section, we will review well known results relating to the transformation of a power spectrum onto the celestial sphere and re-derive the results of~\cite{Raccanelli:2023fle, Raccanelli:2023zkj, Gao:2023rmo}, which constitute a new form of projection which allows for the preservation of information related to the Hyperuranion space-time displacements derived in Sec~\ref{sec:TimeExpansion}.

We begin by noting the usual definition of a projected density field located within a chosen distance window $W(\chi,\chi')$:
\begin{equation}
    \hat{\delta}(\vec{\ell},\chi)=\int\frac{d\chi'}{\chi'^{2}}W(\chi,\chi')\int\frac{dk_{\hat{n}}}{2\pi}e^{-i\chi' k_{\hat{n}}}\delta(k_{\hat{n}},\vec{\ell},\chi')~,
    \label{eq:deltahat}
\end{equation}
where $\delta\left(k_{\hat{n}},\vec{\ell},\chi\right)$ with $\vec{\ell}=\mathbf{k}_{\perp}\chi'$ is the angular field parameterised in a 4D spacetime without accounting for the degeneracy between radial and temporal measurements. 

Taking the correlator of two projected fields as given in Eq.~\eqref{eq:deltahat}, we obtain an expression for the angular power spectrum:
\begin{align}
    \la\hat{\delta}(\vec{\ell}_{1},\chi_{1})\hat{\delta}(\vec{\ell}_{2},\chi_{2})\ra&=\int
    \frac{d\chi'_{1}d\chi'_{2}}{\chi'^{2}_{1}\chi'^{2}_{2}}W(\chi_{1},\chi'_{1})W(\chi_{2},\chi'_{2})\int\frac{dk_{\hat{n},1}}{2\pi}\frac{dk_{\hat{n},2}}{2\pi}e^{-i(\chi'_{1} k_{\hat{n},1}+\chi'_{2}k_{\hat{n},2})}\la\delta(k_{\hat{n},1},\vec{\ell}_{1},\chi_{1})\delta(k_{\hat{n},2},\vec{\ell}_{2},\chi_{2})\ra \nonumber\\
    &=(2\pi)^{3}\hspace{-0.25em}\int
    \frac{d\chi'_{1}d\chi'_{2}}{\chi'^{2}_{1}\chi'^{2}_{2}}W(\chi_{1},\chi'_{1})W(\chi_{2},\chi'_{2})\delta^{2\mathrm{D}}(\tilde{\vec{\ell}}_{1}\hspace{-0.25em}+\tilde{\vec{\ell}}_{2})\int\frac{dk_{\hat{n}}}{2\pi}e^{-ik_{\hat{n}}(\chi'_{1} -\chi'_{2})}\mathcal{P}(k_{\hat{n}},\vec{\ell}_{1},\chi'_{1},\chi'_{2}) \nonumber\\
    &=(2\pi)^{3}\hspace{-0.25em}\int
    \frac{d\chi'_{1}d\chi'_{2}}{\chi'^{2}_{1}\chi'^{2}_{2}}W(\chi_{1},\chi'_{1})W(\chi_{2},\chi'_{2})\delta^{2\mathrm{D}}(\tilde{\vec{\ell}}_{1}\hspace{-0.25em}+\tilde{\vec{\ell}}_{2})\mathbb{C}(\vec{\ell}_{1},\chi'_{1},\chi'_{2})~,
\label{eq:ETCl}    
\end{align}
where $\tilde{\vec{\ell}}=\vec{\ell}/\chi'$, in the second step we have integrated over our momentum delta function, and on the last line we introduced the angular power spectrum
\begin{equation}
    \mathbb{C}(\vec{\ell},\chi_{1},\chi_{2})\equiv\frac{1}{\chi_{1}\chi_{2}}\int \frac{dk_{\hat{n}}}{2\pi}e^{ik_{\hat{n}}(\chi_{1}-\chi_{2})}\mathcal{P}(k_{\hat{n}},\vec{\ell},\chi_{1},\chi_{2})~.
    \label{eq:UTCl1}
\end{equation}

\begin{figure}[h!]
\centering
\includegraphics[width=0.8\textwidth]{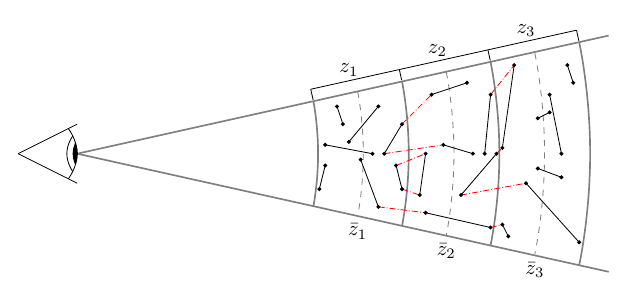}
\caption{
An illustration of a light cone through the celestial sphere showing the sorting of observables, represented by black dots, into a trio of redshift bins and a selection of their correlations. In a standard analysis, every one of these observables would be treated as though it was located at the mean redshift of its bin, shown in the figure by dashed grey lines. The black lines show examples of correlators of pairs of density fields within given bins, while the red dot-dashed lines show examples of cross-bin correlators.
}
\label{fig:RedshiftBins}
\end{figure}

In Fig.~\ref{fig:RedshiftBins} we illustrate the celestial sphere as divided into three redshift bins and the notion of projected power spectra within it. Every observed tracer is located within a given redshift bin. Usually, every field within a given bin would then be treated as being located exactly at the mean redshift of that bin, shown in the figure with dashed grey lines; this is the equal time approximation. In the following section, we will show how a new projection can be defined which allows for the preservation of information relating to the displacement of each tracer field from its respective bin mean.

Cross-bin correlators, shown in the figure as red dashed lines, have been found in most analyses to be substantially smaller in magnitude than correlators assessed within a single bin unless integrated relativistic effects are accounted for; however, the inclusion of such effects leads to them having a similar magnitude to single bin correlators. Furthermore, we note that the radial suppression is magnified by the equal time approximation due to the fact that the two density fields being correlated will be treated as being at their respective bin means; in the event that the fields are located between the means and the border between the two bins, this suppression would be substantially reduced.

\subsection{Correlator-level Projections}
\label{sec:CLP}
We will now briefly review the work first shown in \cite{Raccanelli:2023fle,Raccanelli:2023zkj}, in which power spectra are projected onto the celestial sphere in a manner which allows unequal time terms to manifest which correct power spectra estimated with the equal time approximation.  We begin by noting that, in the case of a predefined correlator, we can treat $\bar{\chi}$ as being the arithmetic mean redshift of the correlator, such that $\delta\chi_{1}=-\delta\chi_{2}=\frac{1}{2}\delta\chi$, where $\delta\chi=\chi_{1}-\chi_{2}$, allowing us to reparametrise Eq.~\eqref{eq:UTCl1} as $\mathbb{C}(\vec{\ell},\bar{\chi},\delta\chi)$.  

We wish to expand upon Eq.~\eqref{eq:ETCl} by incorporating new parameters which will allow for the treatment of information relating to time displacement.  We do his by defining a Fourier counterpart to $\delta\chi$, which we label $q_{\hat{n}}$, and corresponding angular coordinates $\mathbf{q}_{\perp}\equiv\vec{\ell}/\bar{\chi}$, giving us a new three dimensional momentum $\mathbf{q}$ with which we can define 
\begin{equation}
\begin{split}
P(\mathbf{q},\bar{\chi})\equiv& \bar{\chi}^{2}\int d \delta\chi e^{-i\delta\chi q_{\hat{n}}}\mathbb{C}(\vec{\ell},\bar{\chi},\delta\chi)~,\\=&(2\pi)^{3}\int\frac{dk_{\hat{n}}}{2\pi}\int d\delta\chi\epsilon(\bar{\chi},\delta\chi)e^{i\delta\chi (k_{\hat{n}}-q_{\hat{n}})}\mathcal{P}(k_{\hat{n}},\vec{\ell},\bar{\chi})~,
\label{eq:CLCl}
\end{split}
\end{equation}
where we have defined $\epsilon(\chi_{1},\chi_{2})\equiv\bar{\chi}^{2}/(\chi_{1}^{2}\chi_{2}^{2})$. 

Including the time expansion defined in Eq.~\eqref{eq:expandeddelta} for each of the two fields in the power spectrum and likewise Taylor expanding $\epsilon$ up to first order, noting that $\epsilon(\bar{\chi},0)=1$ and that the $n$th term is proportional to $1/\bar{\chi}^{n}\approx 0$, this becomes
\begin{equation}
    P(\mathbf{q},\bar{\chi})=(2\pi)^{3}\int \frac{dk_{\hat{n}}}{2\pi}\int d\delta\chi e^{i\delta\chi(k_{\hat{n}}-q_{\hat{n}})}\left(1+\frac{1}{2}F_{a}(\bar{\chi})H(\bar{\chi})\delta\chi\right)\left(1-\frac{1}{2}F_{a}(\bar{\chi})H(\bar{\chi})\delta\chi\right)\mathcal{P}(k_{\hat{n}},\vec{\ell},\bar{\chi})~.
    \label{eq:expandedCLCl}
\end{equation}

As can be seen, the first order corrections in Eq.~\eqref{eq:expandedCLCl} cancel one another out, leaving only second order and above corrections to account for unequal time effects in the case of an autocorrelator analysis.  However, as shown in~\cite{Raccanelli:2023fle, Raccanelli:2023zkj}, when performing a multi-tracer analysis with differently bias populations, first order corrections survive and can be relevant for future galaxy surveys analyses.

The above analysis applies to a specific correlator with its own mean redshift defined with respect to its constituent tracers.  To generate corrections applicable to a redshift bin, we would wish to integrate over all values of $\bar{\chi}$ that are possible within that bin.  A method to do this by defining a Fourier analogue of the difference between the correlator's mean and the bin's mean, $\delta\bar{\chi}$, is described in App.~\ref{sec:BACL}.

\begin{figure}[h!]
\centering
\includegraphics[width=0.6\textwidth]{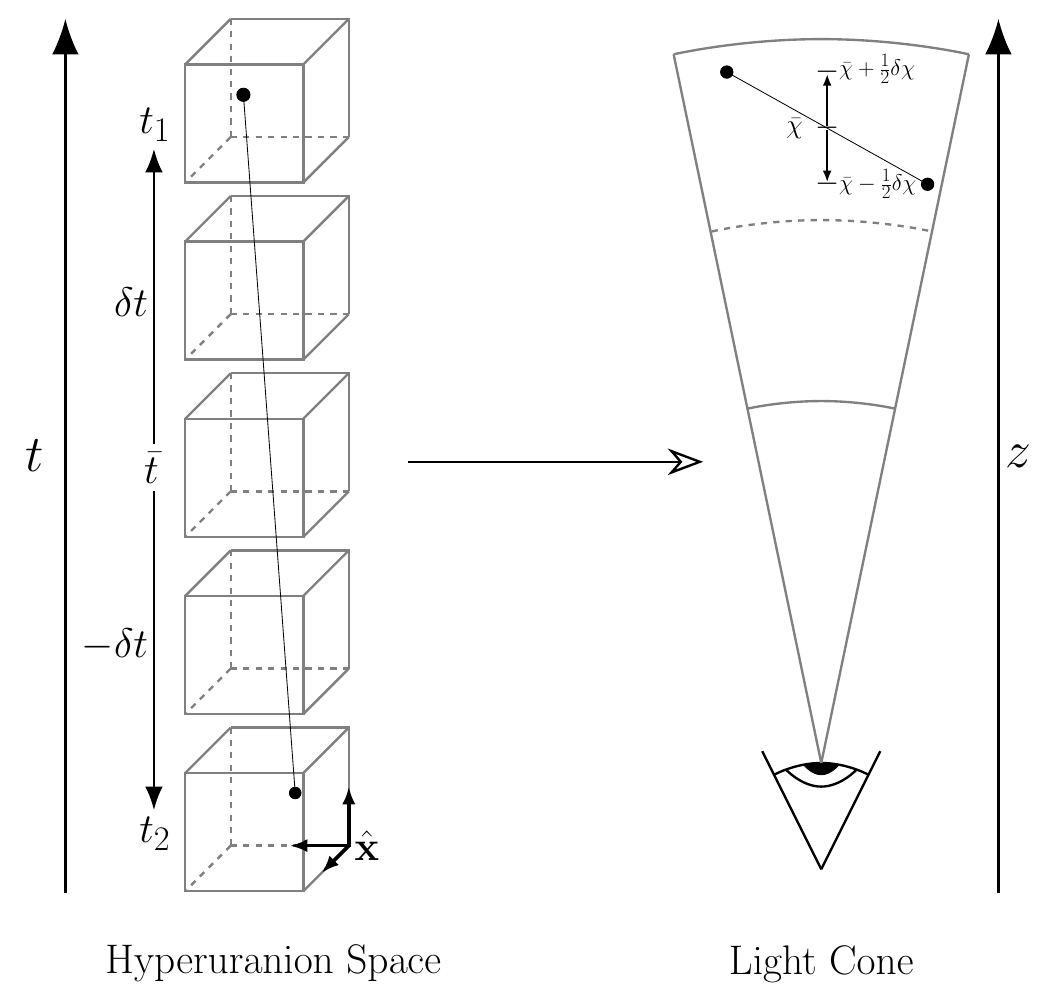}
\caption{An illustration of the correlator-level projection of a power spectrum from Hyperuranion space onto a light cone through the celestial sphere.  In Hyperuranion space, each field is separated by $\pm \delta t$ from the mean time of the power spectrum, $\bar{t}$.  When the power spectrum is projected, this corresponds to each field being located at a comoving distance $\pm \frac{1}{2}\delta\chi$ from the mean comoving distance of the galaxy pair, $\bar{\chi}$. Both fields are located within the same redshift bin, the mean redshift of which is shown as a dotted line on the light cone. In order to generate an averaged power spectrum over the bin, all $\bar{\chi}$ within that bin may be integrated over.}
\label{fig:PtoClCorrelator}
\end{figure}

The correlator-level projection of a power spectrum is illustrated in Fig.~\ref{fig:PtoClCorrelator}. As can be seen, the radial coordinate is translated into the time component in the projection as well as the remaining Cartesian spatial coordinates being converted into angular coordinates. The definition of the $\mathbf{q}$ vector allows for the preservation of information relating to the displacement of the two fields from the average of their two times and its incorporation into the expression for the measured power spectrum. 

\section{Field Level Projections}
\label{sec:FLP}

We now introduce a concept which we refer to as field level projection.  This is a formalism for the projection of density fields onto the celestial sphere followed by the correlation of their projections, in contrast to the correlator-level projection studied in Sec.~\ref{sec:SLP} in which the density fields were correlated in Hyperuranion space and their correlator was then itself projected. Furthermore, unlike in the correlator-level analysis above, we define each density field as being located a given distance $\delta\chi$ from the mean redshift of the bin into which it is sorted, rather than from the mean redshift of a given correlator within that bin.  

We show that the projection of individual fields results in the manifestation of first order correction terms within redshift bins, which account for the difference between the location of each field and the location of the bin's mean.  Furthermore, the allowance for fields to be displaced from separate times before being projected results in new correction terms which account for the evolution of the Universe between those  times and apply to cross-bin analyses.

\subsection{Projecting onto the Celestial Sphere}
\label{sec:PontoCS}
We wish to derive an alternative to Eq.~\eqref{eq:deltahat} which will allow for a time displaced density field to be projected onto the celestial sphere in a manner which preserves an analogue of the information encoded in the corrections derived in Sec.~\ref{sec:TimeExpansion}.  Beginning by Fourier transforming our density field into our $\mathbf{q}$ momentum space defined as in Sec.~\ref{sec:CLP}, we can define a time displaced density field on the celestial sphere as
\begin{equation}
\hat{\delta}\left(\mathbf{q}, \bar \chi\right)\equiv\bar{\chi}^{2}\int d\delta\chi e^{-i\delta\chi q_{\hat{n}}}\hat{\delta}(\mathbf{q}_{\perp},\chi)~,
\label{eq:deltac}
\end{equation}
where $\bar{\chi}$ is the mean redshift of the bin into which the field is sorted.

Inserting Eq.~\eqref{eq:deltahat} into Eq.~\eqref{eq:deltac}, we obtain
\begin{equation}
\hat{\delta}\left(\mathbf{q},\bar{\chi}\right)=\bar{\chi}^{2}\int d\delta\chi\, e^{-i
\delta\chi q_{\hat{n}}}\int\frac{d\chi'}{\chi'^{2}}W(\chi,\chi')\int\frac{dk_{\hat{n}}}{2\pi}e^{-i\chi'k_{\hat{n}}}\delta(k_{\hat{n}},\mathbf{q}_{\perp},\chi')~.
    \label{eq:deltac3}
\end{equation}
At this point we choose to set $W(\chi,\chi')=\delta^{\mathrm{D}}(\chi'-\chi)$.  Thus, we arrive at a final expression for the projected unequal time density field that we will use throughout this section:
\begin{equation}
\hat{\delta}\left(\mathbf{q},\bar{\chi}\right)=\int\frac{dk_{\hat{n}}}{2\pi}e^{-i\bar{\chi}k_{\hat{n}}}\int d\delta\chi \, e^{-i
\delta\chi \left(k_{\hat{n}}+q_{\hat{n}}\right)}\gamma(\chi)\delta(k_{\hat{n}},\mathbf{q}_{\perp},\chi)~,
    \label{eq:deltac4}
\end{equation}
where $\gamma(\chi)=\left(\bar{\chi}/\chi\right)^{2}$.  

Now that we have defined a projection mechanism which will allow Hyperuranion time displacements of individual density fields to be projected onto the celestial sphere, we introduce the results of Sec.~\ref{sec:TimeExpansion} into Eq.~\eqref{eq:deltac4}. We begin by Taylor expanding the time component of the Hyperuranion density field around the mean redshift of the bin into which the field is sorted, obtaining 
\begin{equation}
\hat{\delta}\left(\mathbf{q},\bar{\chi}\right)=\int\frac{dk_{\hat{n}}}{2\pi}e^{-i\bar{\chi}k_{\hat{n}}}\int d\delta\chi e^{-i
\delta\chi \left(k_{\hat{n}}+q_{\hat{n}}\right)}\sum_{j=0}^{\infty}\frac{1}{j!}\frac{d^{j}}{d\chi^{j}}\left(\gamma(\chi)\delta(k_{\hat{n}},\mathbf{q}_{\perp},\chi)\right)\bigg\vert_{\delta\chi=0}(\delta\chi)^{j}~.
\label{eq:deltac5}
\end{equation}
Recognising that
\begin{equation}
    \int d\delta\chi e^{-i\delta\chi\left(k_{\hat{n}}+q_{\hat{n}}\right)}\left(\delta\chi\right)^{j}=2\pi i^{j}\frac{d^{j}}{dq_{\hat{n}}^{j}}\delta^{\mathrm{D}}(k_{\hat{n}}+q_{\hat{n}})~,
    \label{eq:deltastep}
\end{equation}
this becomes
\begin{equation}
\hat{\delta}\left(\mathbf{q},\bar{\chi}\right)=\sum_{j=0}^{\infty}\frac{i^{j}}{j!}\frac{d^{j}}{dq_{\hat{n}}^{j}}\left[e^{-i\bar{\chi}q_{\hat{n}}}\frac{d^{j}}{d\chi^{j}}\left(\gamma(\chi)\delta(\mathbf{q},\chi)\right)\bigg\vert_{\delta\chi=0}\right]~.
\label{eq:deltac4a}
\end{equation}
The derivatives of $\gamma(\chi)$ in this limit are $\partial_{\chi}^{j}\gamma(\chi)=(-1)^{j}(1+j)!/\bar{\chi}^{j}\approx 0$.  Inserting Eq.~\eqref{eq:expandeddelta} into Eq.~\eqref{eq:deltac4a}, we now have
\begin{equation}
\hat{\delta}\left(\mathbf{q},\bar{\chi}\right)=
\left[1+iF_{a}(\bar{\chi})H(\bar{\chi})\partial_{q_{\hat{n}}}\right]e^{-i\bar{\chi}q_{\hat{n}}}\delta(\mathbf{q},\bar{\chi})~.
\label{eq:deltac5a}
\end{equation}
To generalise this to the case of biased tracers, we include biasing parameters by inserting Eq.~\eqref{eq:biasedexpansion} into Eq.~\eqref{eq:deltac4}; thus, the biased projected field can be written as
\begin{equation}
    \hat{\delta}(\mathbf{q},\bar{\chi})= \left[b(\bar{\chi})+i\left(F_{a}(\bar{\chi})b(\bar{\chi})+\partial_{z}b(\bar{\chi})\right)H(\bar{\chi})\partial_{q_{\hat{n}}}\right]e^{-i\bar{\chi}q_{\hat{n}}}\delta(\mathbf{q},\bar{\chi})~.
    \label{eq:deltac6}
\end{equation}
Eqs.~\eqref{eq:deltac5a} and \eqref{eq:deltac6} constitute expressions for an individual density contrast field projected onto the celestial sphere in a manner which preserves a displacement from a predefined redshift, which we are taking to be the mean redshift of the bin in which the field is measured.

We note that the phase term remaining from the Fourier transforms is operated on by the derivative operator.  As will be seen in the following study of the power spectrum, this results in new correction terms which apply to cross-bin correlators.

We will now develop two methods for applying the field level projection to power spectra.  One of these methods will consist of applying the transformations defined in Eq.~\eqref{eq:deltac} to a Hyperuranion power spectrum, in effect, projecting the two fields of a pre-defined power spectrum independently.  The other will consist of displacing the fields individually as done in Eq.~\eqref{eq:expandeddelta}, applying the Fourier transform defined in Eq.~\eqref{eq:deltac}, and then taking the expectation value of the product of a pair of the ensuing projected fields.  We show that both methods generate the same results and numerically analyse the results in the context of a biased single tracer power spectrum.

\subsubsection{Displaced Correlators}
\label{sec:SLP}
Transforming a power spectrum with two instances of the operation given in Eq.~\eqref{eq:deltac4} gives us
\begin{equation}
\begin{split}
\la\hat{\delta}\left(\mathbf{q}_{1},\bar{\chi}_{1}\right)\hat{\delta}\left(\mathbf{q}_{2},\bar{\chi}_{2}\right)\ra\hspace{-0.05cm}=&\hspace{0.05cm}(2\pi)^{3}\hspace{-0.075cm}\int\frac{dk_{\hat{n},1}}{2\pi}\int\frac{dk_{\hat{n},2}}{2\pi}e^{-i(\bar{\chi}_{1}k_{\hat{n},1}+\bar{\chi}_{2}k_{\hat{n},2})}\hspace{-0.1cm}\int d\delta\chi_{1}\int d\delta\chi_{2} e^{-i\left(\delta\chi_{1} \left(k_{\hat{n},1}+q_{\hat{n},1}\right)+\delta\chi_{2}\left(k_{\hat{n},2}+q_{\hat{n},2}\right)\right)}\\&\times\gamma(\chi_{1})\gamma(\chi_{2})\delta^{3\mathrm{D}}\hspace{-0.125cm}\left(\mathbf{k}_{1}+\mathbf{k}_{2}\right)\mathcal{P}(\mathbf{k}_{1},\mathbf{k}_{2},\chi_{1},\chi_{2}).
    \label{eq:specdeltac1}
    \end{split}
\end{equation}
Implementing Eq.~\eqref{eq:deltac5} into Eq.~\eqref{eq:specdeltac1}, we obtain
\begin{equation}
\begin{split}
\la\hat{\delta}\left(\mathbf{q}_{1},\bar{\chi}_{1}\right)\hat{\delta}\left(\mathbf{q}_{2},\bar{\chi}_{2}\right)\ra=&\hspace{0.05cm}(2\pi)^{3}\hspace{-0.075cm}\int\frac{dk_{\hat{n},1}}{2\pi}\int\frac{dk_{\hat{n},2}}{2\pi}e^{-i(\bar{\chi}_{1}k_{\hat{n},1}+\bar{\chi}_{2}k_{\hat{n},2})}\hspace{-0.1cm}\int d\delta\chi_{1}\int d\delta\chi_{2} e^{-i\left(\delta\chi_{1} \left(k_{\hat{n},1}+q_{\hat{n},1}\right)+\delta\chi_{2}\left(k_{\hat{n},2}+q_{\hat{n},2}\right)\right)}\\&\times\left[1+F_{a}(\bar{\chi}_{1})H(\bar{\chi}_{1})\delta\chi_{1}\right]\left[1+F_{a}(\bar{\chi}_{2})H(\bar{\chi}_{2})\delta\chi_{2}\right]\delta^{3\mathrm{D}}\hspace{-0.125cm}\left(\mathbf{k}_{1}+\mathbf{k}_{2}\right)\mathcal{P}(\mathbf{k}_{1},\mathbf{k}_{2},\bar{\chi}_{1},\bar{\chi}_{2})~.
    \label{eq:specdeltac2}
    \end{split}
\end{equation}

Expanding upon Eq.~\eqref{eq:deltastep} by recognising that
\begin{equation}
    \int \frac{dk_{\hat{n}}}{2\pi}e^{-i\bar{\chi}k_{\hat{n}}}\delta^{3\mathrm{D}}\hspace{-0.125cm}\left(\mathbf{k}_{1}+\mathbf{k}_{2}\right)\int d\delta\chi e^{-i\delta\chi\left(k_{\hat{n}}+q_{\hat{n}}\right)} \left(\delta\chi\right)^{j}=i^{j}\frac{d^{j}}{dq_{\hat{n}}^{j}}e^{-i\bar{\chi}q_{\hat{n}}}\delta^{3\mathrm{D}}\hspace{-0.125cm}\left(\mathbf{q}_{1}+\mathbf{q}_{2}\right)~,
\end{equation}
Eq.~\eqref{eq:specdeltac2} becomes
\begin{equation}
\begin{split}
\la\hat{\delta}\left(\mathbf{q}_{1},\bar{\chi}_{1}\right)\hat{\delta}\left(\mathbf{q}_{2},\bar{\chi}_{2}\right)\ra=&(2\pi)^{3}\left[1+iF_{a}(\bar{\chi}_{1})H(\bar{\chi}_{1})\partial_{q_{\hat{n},1}}\right]\left[1+iF_{a}(\bar{\chi}_{2})H(\bar{\chi}_{2})\partial_{q_{\hat{n},2}}\right]e^{-i\left(\bar{\chi}_{1}q_{\hat{n},1}+\bar{\chi}_{2}q_{\hat{n},2}\right)}\delta^{3\mathrm{D}}\hspace{-0.125cm}\left(\mathbf{q}_{1}+\mathbf{q}_{2}\right)\\&\times\mathcal{P}(\mathbf{q}_{1},\mathbf{q}_{2},\bar{\chi}_{1},\bar{\chi}_{2})~.
\label{eq:specdeltacfinal}
\end{split}
\end{equation}
 In order to generalise our results to biased tracers, we will now incorporate a linear biasing parameter into each field.  Using Eq.~\eqref{eq:biasexp} and repeating the above analysis, the biased generalisation of Eq.~\eqref{eq:specdeltacfinal} can be seen to be
\begin{equation}
\begin{split}
\la\hat{\delta}\left(\mathbf{q}_{1},\bar{\chi}_{1}\right)\hat{\delta}\left(\mathbf{q}_{2},\bar{\chi}_{2}\right)\ra=(2\pi)^{3}&\left[b_{1}(\bar{\chi}_{1})+i\left(F_{a}(\bar{\chi}_{1})b_{1}(\bar{\chi}_{1})+\partial_{z}b_{1}(\bar{\chi}_{1})\right)H(\bar{\chi}_{1})\partial_{q_{\hat{n},1}}\right]\\\times&\left[b_{2}(\bar{\chi}_{2})+i\left(F_{a}(\bar{\chi}_{2})b_{2}(\bar{\chi}_{2})+\partial_{z}b_{2}(\bar{\chi}_{2})\right)H(\bar{\chi}_{2})\partial_{q_{\hat{n},2}}\right]\\\times& e^{-i(\bar{\chi}_{1}q_{\hat{n},1}+\bar{\chi}_{2}q_{\hat{n},2})}\delta^{3\mathrm{D}}\hspace{-0.125cm}\left(\mathbf{q}_{1}+\mathbf{q}_{2}\right)\mathcal{P}(\mathbf{q}_{1},\mathbf{q}_{2},\bar{\chi}_{1},\bar{\chi}_{2})~.
\label{eq:SLcorrelators}
\end{split}
\end{equation}

\subsubsection{Displaced Fields}
We will now rederive the above equations from the point of view of projecting the fields individually onto the celestial sphere and only then correlating them.  We do this by taking the expectation value of the product of two individually projected fields, each as described in Eq.~\eqref{eq:deltac5}.  Up to first order in the time expansion, this gives us
\begin{equation}
\begin{split}
\la\hat{\delta}\left(\mathbf{q}_{1},\bar{\chi}_{1}\right)\hat{\delta}\left(\mathbf{q}_{2},\bar{\chi}_{2}\right)\ra=&\left[1+iF_{a}(\bar{\chi}_{1})H(\bar{\chi}_{1})\partial_{q_{\hat{n},1}}\right]\left[1+iF_{a}(\bar{\chi}_{2})H(\bar{\chi}_{2})\partial_{q_{\hat{n},2}}\right]e^{-i\left(\bar{\chi}_{1}q_{\hat{n},1}+\bar{\chi}_{2}q_{\hat{n},2}\right)}\la\delta\left(\mathbf{q}_{1},\bar{\chi}_{1}\right)\delta\left(\mathbf{q}_{2},\bar{\chi}_{2}\right)\ra \\=&(2\pi)^{3}\left[1+iF_{a}(\bar{\chi}_{1})H(\bar{\chi}_{1})\partial_{q_{\hat{n},1}}\right]\left[1+iF_{a}(\bar{\chi}_{2})H(\bar{\chi}_{2})\partial_{q_{\hat{n},2}}\right]e^{-i\left(\bar{\chi}_{1}q_{\hat{n},1}+\bar{\chi}_{2}q_{\hat{n},2}\right)}\\&\times\delta^{3\mathrm{D}}\hspace{-0.125cm}\left(\mathbf{q}_{1}+\mathbf{q}_{2}\right)\mathcal{P}_{n}(\mathbf{q}_{1},\mathbf{q}_{2},\bar{\chi}_{1},\bar{\chi}_{2})~.
\label{eq:correlators}
\end{split}
\end{equation}
Including biasing and repeating the derivation, we see that
\begin{equation}
    \begin{split}
\la\hat{\delta}\left(\mathbf{q}_{1},\bar{\chi}_{1}\right)\hat{\delta}\left(\mathbf{q}_{2},\bar{\chi}_{2}\right)\ra
=(2\pi)^{3}&\left[b_{1}(\bar{\chi}_{1})+i\left(F_{a}(\bar{\chi}_{1})b_{1}(\bar{\chi}_{1})+\partial_{z}b_{1}(\bar{\chi}_{1})\right)H(\bar{\chi}_{1})\partial_{q_{\hat{n},1}}\right]\\\times&\left[b_{2}(\bar{\chi}_{2})+i\left(F_{a}(\bar{\chi}_{2})b_{2}(\bar{\chi}_{2})+\partial_{z}b_{2}(\bar{\chi}_{2})\right)H(\bar{\chi}_{2})\partial_{q_{\hat{n},2}}\right]\\\times& e^{-i(\bar{\chi}_{1}q_{\hat{n},1}+\bar{\chi}_{2}q_{\hat{n},2})}\delta^{3\mathrm{D}}\hspace{-0.125cm}\left(\mathbf{q}_{1}+\mathbf{q}_{2}\right)\mathcal{P}(\mathbf{q}_{1},\mathbf{q}_{2},\bar{\chi}_{1},\bar{\chi}_{2})~.
\label{eq:biasutspec}
\end{split}
\end{equation}
We note that Eq.~\eqref{eq:correlators} is the same as Eq.~\eqref{eq:specdeltacfinal} and Eq.~\eqref{eq:biasutspec} is the same as Eq.~\eqref{eq:SLcorrelators}, thus confirming that the two approaches to unequal time field level projections give the same results.

\subsubsection{Summary}

Having defined the two point unequal time ensemble average in momentum space and shown that, using the field level projection formalism shown in Eq.~\eqref{eq:deltac5}, the operations of correlation and projection commute, we may now note that the field level projection formalism places the phase terms and the delta function within the domain of the derivative operator. Taking these derivatives yields correction terms which were not present in the correlator-level analysis as previously defined. 
 The magnitude of these corrections in the context of a linearly biased, single tracer power spectrum is the subject of the next section.  

In App.~\ref{sec:BACL}, we show that with a suitable bin averaging scheme based around a Fourier transform of the comoving distance between the bin's mean redshift and the mean redshift's of its contained correlators, these corrections can be replicated at least in the context of a single bin analysis.

These corrections arise from the introduction of two Fourier transforms.  This allows the two fields to be independently displaced from the reference redshift, taken in this analysis to be the mean of the bin into which the field is sorted.  This removes the assumption that $\delta\chi_{1}=-\delta\chi_{2}$ which leads to the cancellation of first order corrections in the unextended correlator-level projection.

\begin{figure}[h!]
\centering
\includegraphics[width=0.6\textwidth]{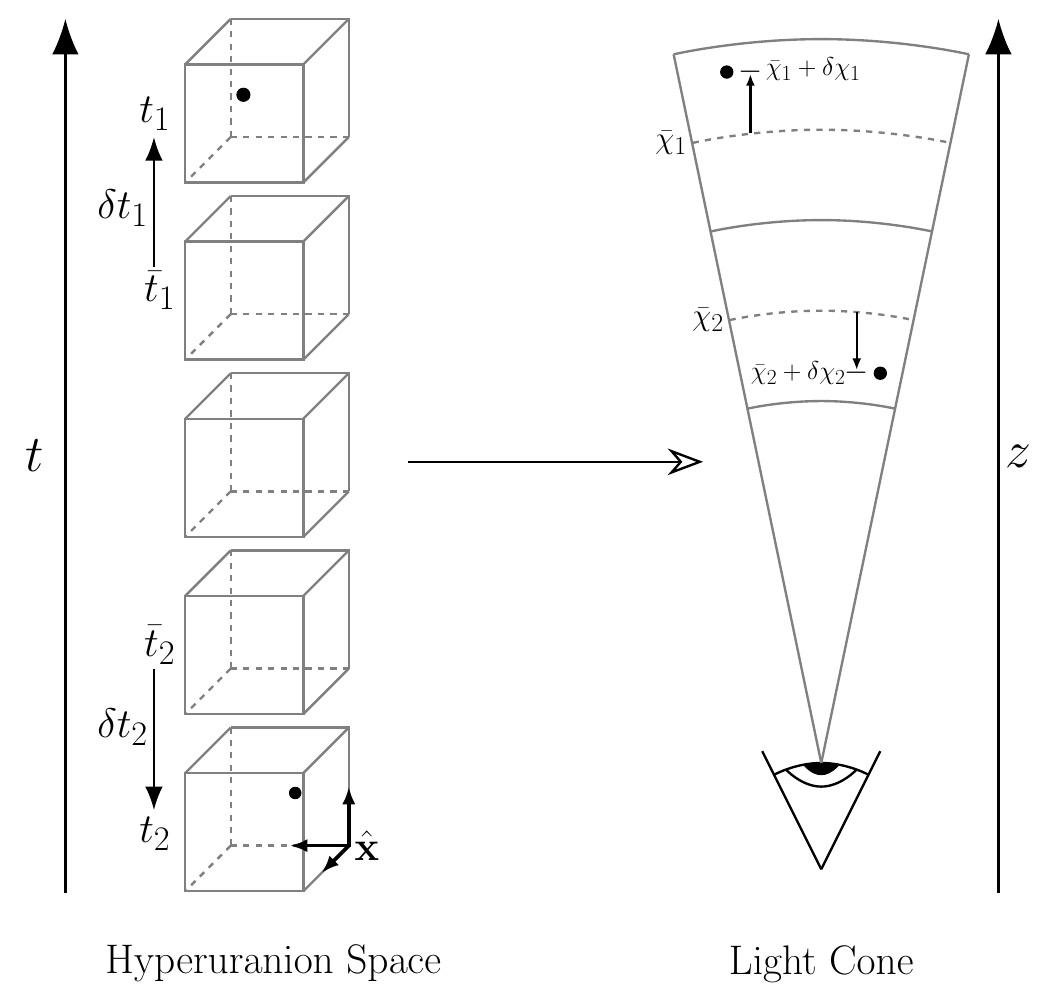}
\caption{An illustration of the field level projection of a pair of fields from Hyperuranion space onto 
a light cone through the celestial sphere.  In Hyperuranion space, each field is separated by a given $\delta t$ from its assumed time $\bar{t}$.  When projected, this corresponds to each field being defined at a distance $\delta\chi$ from the mean comoving distance of its redshift bin, $\bar{\chi}$.  After having been projected, the fields may then be correlated.  
}
\label{fig:PtoClField}
\end{figure}

The field level projection of a pair of density fields in preparation for their correlation into an unequal time power spectrum is illustrated in Fig.~\ref{fig:PtoClField}. The unique times of the two fields are defined with respect to assumed times corresponding to the mean redshifts of the bins into which they will be sorted.  We note that we can set $\bar{\chi}_{1}=\bar{\chi}_{2}$ to sort them both into the same bin or treat them in the context of cross-bin analyses. 

\section{Evaluating The Power Spectrum}
\label{sec:pk}
Taking Eq.~\eqref{eq:correlators} and integrating over $\mathbf{q}_{2}$, we obtain 
\begin{equation}
\begin{split}
\label{eq:intPc}
\int d\mathbf{q}_{2}\la \hat{\delta}(\mathbf{q}_{1}, \bar \chi_{1}) \hat{\delta}(\mathbf{q}_{2}, \bar \chi_{2}) \ra
=&(2\pi)^{3} \, e^{i (\bar \chi_{2}-\bar{\chi}_{1})q_{\hat{n},1}}\left[1+F_{a}(\bar{\chi}_{1})H(\bar{\chi}_{1})\left(\bar{\chi}_{1}-\bar{\chi}_{2}+i\partial_{q_{\hat{n},1}}\right)\right]\mathcal{P}(\mathbf{q}_{1},\bar{\chi}_{1},\bar{\chi}_{2})\\ =&(2\pi)^{3}\, e^{i (\bar \chi_{2}-\bar{\chi}_{1})q_{\hat{n},1}}\left[1+R(\bar{\chi}_{1},\bar{\chi}_{2})+Y(\mathbf{q}_{1},\bar{\chi}_{1},\bar{\chi}_{2})\right]\mathcal{P}(\mathbf{q}_{1},\bar{\chi}_{1},\bar{\chi}_{2})~,
\end{split}
\end{equation}
where we have defined the phase independent cross-bin correction function $R$ and the phase independent within-bin correction function $Y$
\begin{align}
    R(\bar{\chi}_{1},\bar{\chi}_{2})\equiv& ~F_{a}(\bar{\chi}_{1})H(\bar{\chi}_{1})(\bar{\chi}_{1}-\bar{\chi}_{2})~,\\
    Y(\mathbf{q},\bar{\chi}_{1},\bar{\chi}_{2})\equiv& ~iF_{a}(\bar{\chi}_{1})H(\bar{\chi}_{1})\frac{\partial_{q_{\hat{n}}}\mathcal{P}(\mathbf{q},\bar{\chi}_{1},\bar{\chi}_{2})}{\mathcal{P}(\mathbf{q},\bar{\chi}_{1},\bar{\chi}_{2})}
    \label{eq:W2}
\end{align}
and can now see the unequal time power spectrum to be
\begin{equation}
    P(\mathbf{q},\bar{\chi}_{1},\bar{\chi}_{2})=e^{i (\bar \chi_{2}-\bar{\chi}_{1})q_{\hat{n}}}\left[1+R(\bar{\chi}_{1},\bar{\chi}_{2})+Y(\mathbf{q},\bar{\chi}_{1},\bar{\chi}_{2})\right]\mathcal{P}(\mathbf{q},\bar{\chi}_{1},\bar{\chi}_{2})~.
\end{equation}
It is important to note that $R(\bar{\chi}_{1},\bar{\chi}_{2})\neq R(\bar{\chi}_{2},\bar{\chi}_{1})$ and $Y(\mathbf{q},\bar{\chi}_{1},\bar{\chi}_{2})\neq Y(\mathbf{q},\bar{\chi}_{2},\bar{\chi}_{1})$, following from the dependence in Eq.~\eqref{eq:intPc} on the choice of which momentum to integrate over.  This dependence originates with the derivative of the delta function in Eq.~\eqref{eq:correlators}, which negates the derivative of the exponential in the case where the derivative is with respect to the same momentum being integrated over and generates a new derivative of the phase term otherwise.

$R$ originates with the integral over the derivative of the phase term and delta function and accounts for the evolution of the Universe between the two assumed redshifts the fields have been displaced from.  $Y$ originates with the integral over the derivatives of the delta function and the power spectrum and accounts for the displacements themselves; specifically, it accounts for integrated effects of time displacement along the length of the bin.  

Taking a toy model power spectrum, we calculate values of $Y$ as a function of $q_{\hat{n}}$ for $\bar{z}=1$ and plot them in the top left panel of Fig.~\ref{fig:correction}. As can be seen, the correction becomes more pronounced the smaller the value of $q_{\hat{n}}$, which corresponds to larger physical distances between the fields and the bin mean redshift, as one could expect. The dependence of $Y$ upon the redshift of the bin is small, as the term accounts for deviations of the field's redshift from a bin's mean and so is a function of the bin's width more than of its mean value.

The $q_{\hat{n}}$ parameter represents deviation from the bin mean redshift.  As such, in order to generate an overall correction for a redshift bin, we may wish to integrate over the values that it can take.  To this end, we define
 \begin{equation}
\begin{split}
    \mathrm{P}(\mathbf{q}_{\perp},\bar{\chi}_{1},\bar{\chi}_{2})&\equiv\int dq_{\hat{n}}P(\mathbf{q},\bar{\chi}_{1},\bar{\chi}_{2})\\&=\int dq_{\hat{n}}e^{i (\bar \chi_{2}-\bar{\chi}_{1})q_{\hat{n}}}\left[1+R(\bar{\chi}_{1},\bar{\chi}_{2})+Y(\mathbf{q},\bar{\chi}_{1},\bar{\chi}_{2})\right]\mathcal{P}(\mathbf{q},\bar{\chi}_{1},\bar{\chi}_{2})~.
    \label{eq:intq1}
\end{split} 
 \end{equation}
We note that this takes the form of an inverse Fourier transform from $q_{\hat{n}}$ to the distance between the mean redshifts of the bins in question and that in the case of a single bin analysis it simplifies to 
\begin{equation}
    \mathrm{P}(\mathbf{q}_{\perp},\bar{\chi})=\int dq_{\hat{n}}\left[1+Y(\mathbf{q},\bar{\chi})\right]\mathcal{P}(\mathbf{q},\bar{\chi})~.
    \label{eq:intq2}
\end{equation}
The magnitude of the corrections introduced in Eq.~\eqref{eq:intq2} to a single bin power spectrum centred at $z=1$ is shown in the top right panel of Fig.~\ref{fig:correction}.  As can be seen, the integrated imaginary correction becomes percent level with respect to the uncorrected, real integrated power spectrum when the width of the bin is of order $\mathcal{O}(100\mathrm{Mpc~h}^{-1}) $, which is within the size range of many upcoming survey redshift bins, and the modes are studied up to those with wavelengths of the order of $\mathcal{O}(10\mathrm{Mpc~h}^{-1})$.  

The procedure shown in Eq.~\eqref{eq:intPc} can be interpreted as measuring the correlation of a mode for one field with every mode of the other.  Due to the delta function in Eq.~\eqref{eq:CorrSpec}, this would simply be equal to the correlation of the mode with a mode of equivalent wave number when studied in the context of the equal time approximation.  As such, the new correction terms can be seen as having arisen from off-diagonal elements of the delta function which allow for cross-mode correlation.  The procedure shown in Eq.~\eqref{eq:intq1} then consists of integrating over all of the modes of that first field, generating the equivalent of a power spectrum integrated over all modes within a redshift bin but in the context of time displaced density fields.
 
\subsubsection{Biased Tracers}
Generalising Eq.~\eqref{eq:intPc} by repeating the derivation in the context of biased tracers, we see that
\begin{equation}
\begin{split}
\label{eq:biasedintPc}
\int d\mathbf{q}_{2}\la \hat{\delta}(\mathbf{q}_{1}, \bar \chi_{1})\hat{\delta}(\mathbf{q}_{2}, \bar \chi_{2}) \ra
=&(2\pi)^{3}  e^{i (\bar \chi_{2}-\bar{\chi}_{1})q_{\hat{n},1}}b_{2}(\bar{\chi}_{2})\left[b_{1}(\bar{\chi}_{1})+\left(F_{a}(\bar{\chi}_{1})b_{1}(\bar{\chi}_{1})+\partial_{z}b_{1}(\bar{\chi}_{1})\right)H(\bar{\chi}_{1})\left(\bar{\chi}_{1}-\bar{\chi}_{2}\right)\right.\\&\left.+ib_{1}(\bar{\chi}_{1})F_{a}(\bar{\chi}_{1})H(\bar{\chi}_{1})\partial_{q_{\hat{n},1}}\right]\mathcal{P}(\mathbf{q}_{1},\bar{\chi}_{1},\bar{\chi}_{2})\\=&(2\pi)^{3}e^{i(\bar{\chi}_{2}-\bar{\chi}_{1})q_{\hat{n},1}}b_{1}(\bar{\chi}_{1})b_{2}(\bar{\chi}_{2})\left[1+R(\bar{\chi}_{1},\bar{\chi}_{2})+Y(\mathbf{q}_{1},\bar{\chi}_{1},\bar{\chi}_{2})\right]\mathcal{P}(\mathbf{q}_{1},\bar{\chi}_{1},\bar{\chi}_{2})~.
\end{split}
\end{equation}
where
\begin{equation}
    R(\bar{\chi}_{1},\bar{\chi}_{2})=\left[F_{a}(\bar{\chi}_{1})+\frac{\partial_{z}b_{1}(\bar{\chi}_{1})}{b_{1}(\bar{\chi}_{1})}\right]H(\bar{\chi}_{1})\left(\bar{\chi}_{1}-\bar{\chi}_{2}\right)
    \label{eq:R2}
\end{equation}
and $Y$ is defined as in Eq.~\eqref{eq:W2}.  We note the introduction of a new correction term which we have incorporated into $R$, which accounts for the evolution of biasing between the two reference times.  These new terms can be interpreted as the evolution of the biasing between a given tracer and the underlying matter field between the mean times of two redshift bins.  

\begin{figure}[h!]
\centering
\includegraphics[width=0.49\linewidth]{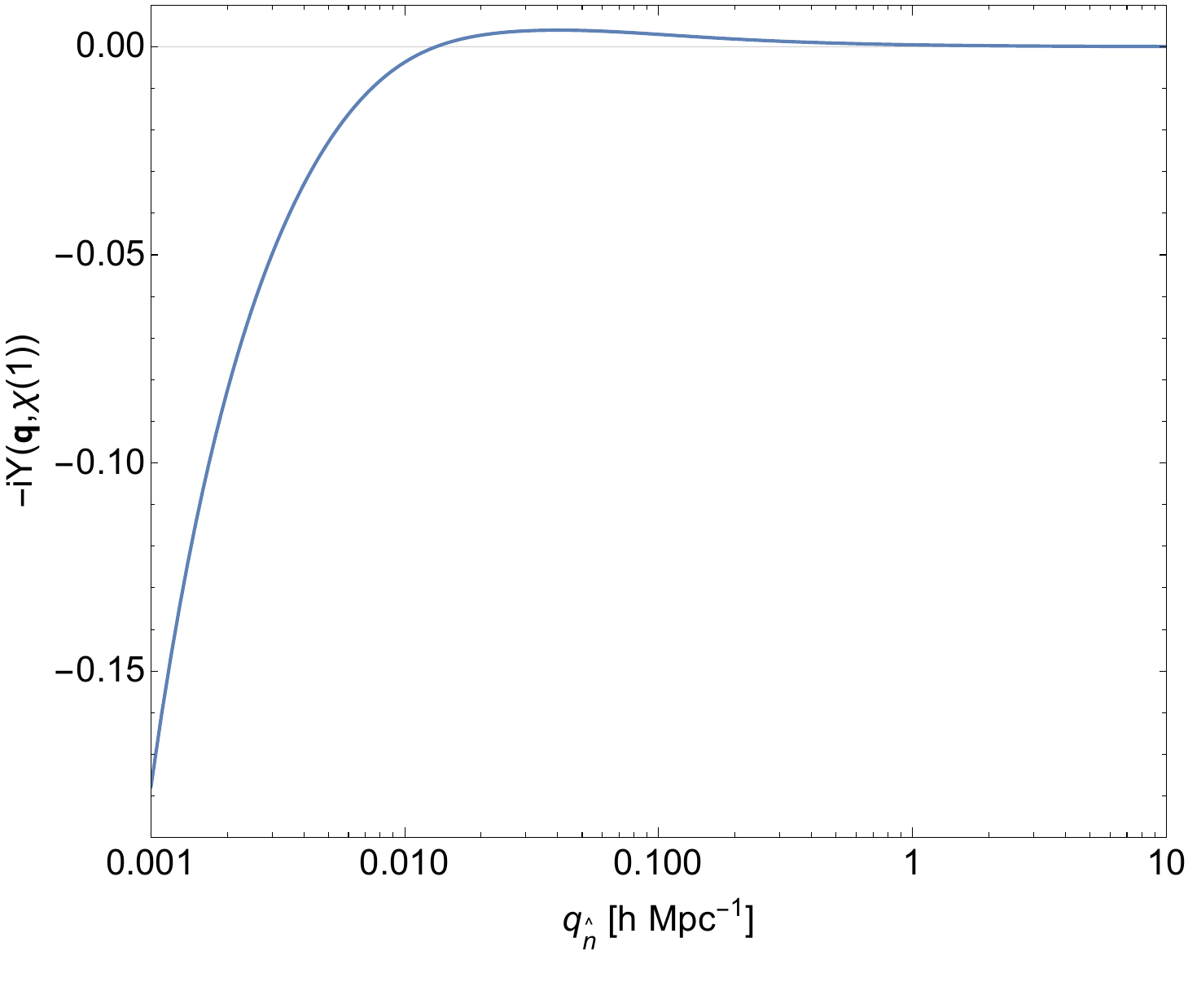}
\includegraphics[width=0.495\linewidth]{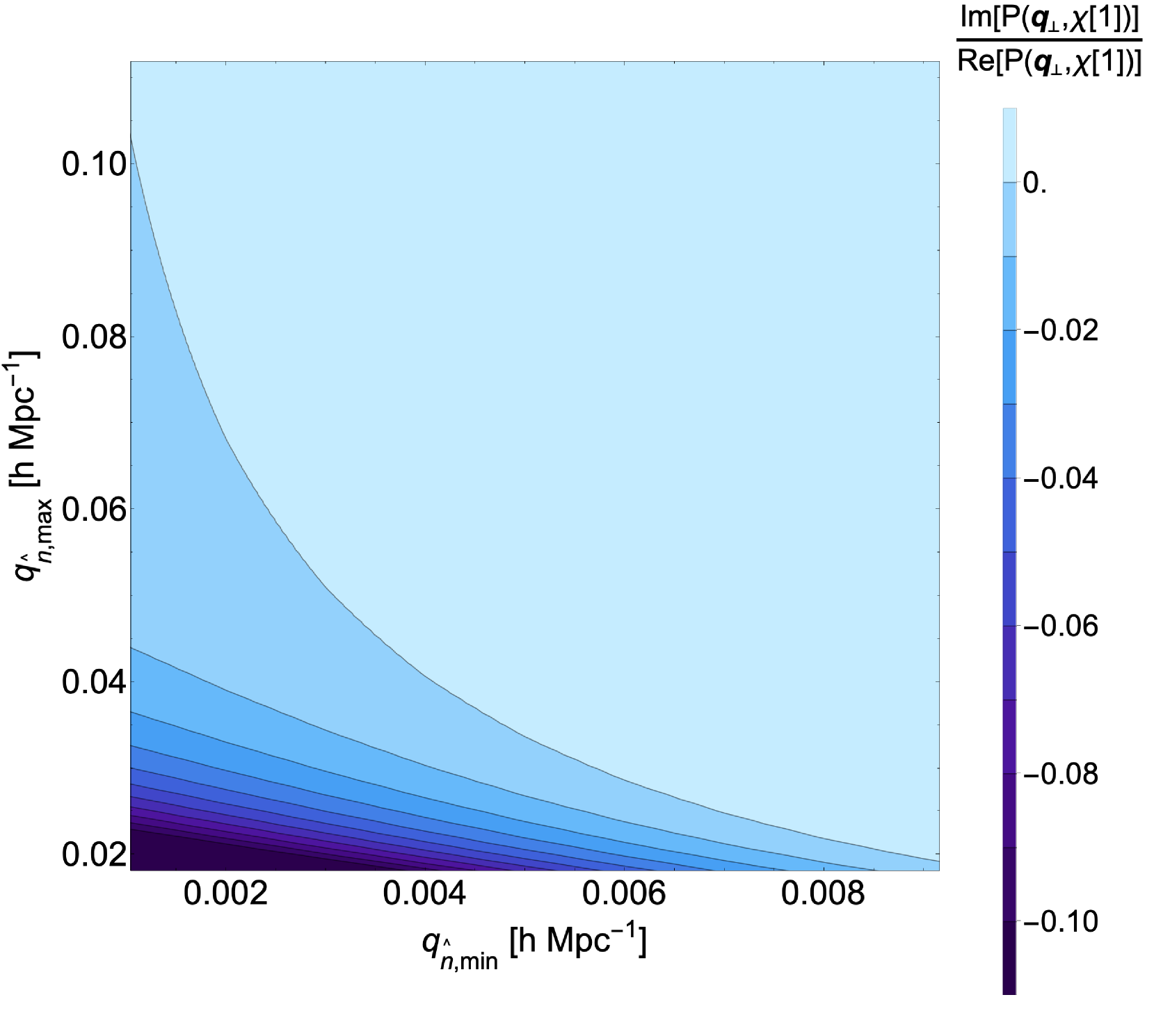}
\includegraphics[width=0.4965\linewidth]{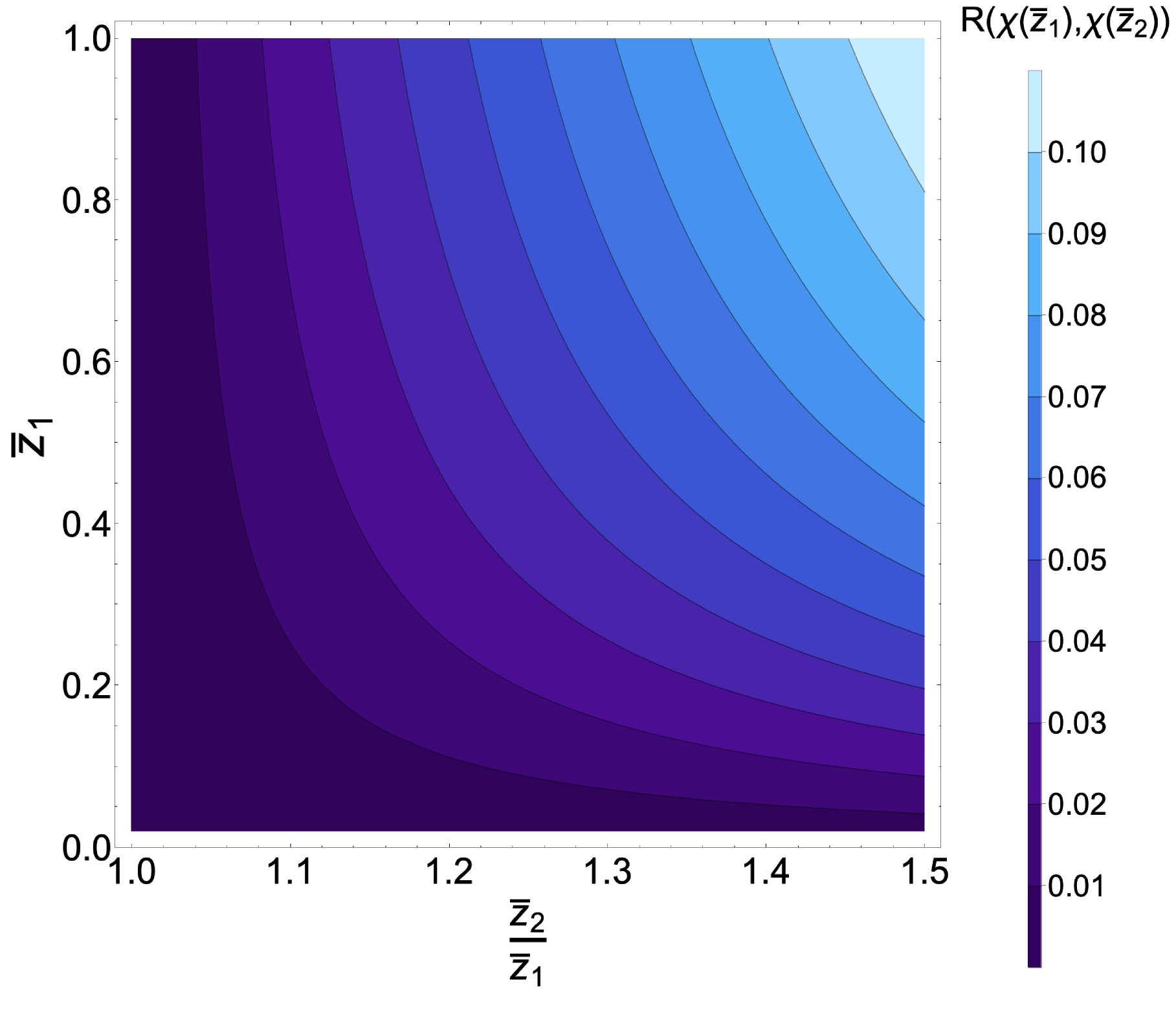}
\caption{\textit{Top left panel:} The magnitude of the within-bin correction factor $Y$ as given by applying Eq.~\eqref{eq:W2} to a toy model power spectrum using the BBKS function.  As can be seen, the corrections become percent level when deviations from the bin mean exceed $q_{\hat{n}}\approx 10^{-2}\mathrm{h~Mpc}^{-1}$.  We have set $\bar{z}=1$, $\Omega_{\mathrm{m}}=0.272$, $h=0.67$, and $k_{\mathrm{nl}}=0.3 \mathrm{h~Mpc^{-1}}$.  \textit{Top right panel:} The ratio of the imaginary correction term in Eq.~\eqref{eq:intq2} with the integrated, uncorrected power spectrum term from the same equation.  These calculations were done with a toy model power spectrum using the BBKS function at bin mean redshift $\bar{z}=1$ and the axes show the limits of the integration, which correspond to the longest and shortest modes viable within a given redshift bin.  In the case of a bin with a width of $\mathcal{O}(100\mathrm{Mpc~h^{-1}})$, the corrections become percent level with respect to the real power spectrum when the largest modes studied correspond to scales of $\mathcal{O}(10\mathrm{Mpc~h^{-1}})$.  We notice that the integrated values of $Y$ vanish as the integrals limits diverge; this is in keeping with the notion that the two momentum integrals commute, which must result in the loss of both field's single bin correction terms.  \textit{Bottom panel:} The cross-bin correction factor $R$ for the biased power spectrum as given in Eq.~\eqref{eq:R2} with $b(\chi)=\sqrt{1+z(\chi)}$. As can be seen, the corrections reach percent level even for closely spaced bins.  }
\label{fig:correction}
\end{figure}

Furthermore, we note that a full integral over $Y$ should result in the vanishing of the remaining single bin corrections.  This can be established by noting that, for physical reasons, the integrals over the two momenta $q_{\hat{n}}$ must commute, such that the loss of one of the field's correction terms by the integral over the other field's momentum indicates a loss of \textit{that} correction term with an integral over the other.  We also note that the integral over the delta function was taken in the above analysis to have infinite limits, in accordance with the usual definition of a Fourier transform.  This fact does allow for a certain asymmetry in as much as an integral with different limits over the remaining momentum would be expected to retain some corrections, as the operations over the two momenta are not the same and so should not be expected to commute.  This is shown in Fig.~\ref{fig:correction}, where we see that an integral over only a part of the momentum's space retains a correction while the full integral with divergent limits does not.

Taking as an example a single tracer analysis with biasing given by $b(\chi)=\sqrt{1+z(\chi)}$, we calculate $R$ as given in Eq.~\eqref{eq:R2} over a range of redshifts and obtain the results shown in the bottom panel of Fig.~\ref{fig:correction}. As can be seen, the corrections increase with redshift spacing and are substantial even with closely spaced bins; however, accounting for the suppressed nature of cross-bin spectra, the importance of these terms will remain uncertain until integrated relativistic effects are accounted for in a future paper.  

A detailed analysis for the magnitude of such corrections for different models of the bias including multi-tracer correlators  will be presented in a follow up work.

\section{Discussion}
\label{sec:discussion}

The statistical analysis of LSS traditionally consists of modeling the evolution of density fields and their correlators in a 4D universe which we refer to as Hyperuranion space and, in the case of angular correlators, projecting them onto the celestial sphere, in doing so removing the time dimension. In many cases, the analysis of LSS data has consisted of sorting observables into redshift bins, treating every tracer in a bin as being located exactly at the effective mean redshift of that bin, and measuring their correlators in the context of this equal time approximation.
This leads to errors that have however traditionally been assumed to be below the precision limitations of the surveys in question. In the era of precision cosmology, it is becoming increasingly important to test such approximations, estimate the errors they cause, and correct for those errors if they are large enough to affect the results of an analysis. 

A method intended to address this issue has existed in the literature for several years~\cite{Scoccimarro:2015bla} in the context of accounting for RSD effects, which locally expands around a projected power spectrum using a Fourier transform of the difference between the redshifts of the tracers. This formalism consists of defining an unequal Fourier space term, assumed to encompass all unequal time effects without recourse to a Hyperuranion correction, and applying it to a projected correlator.  

In~\cite{Raccanelli:2023fle, Raccanelli:2023zkj, Gao:2023tcd}, a new formalism was derived to both quantify and correct for the errors caused by the equal time approximation, which we refer to as correlator-level projection. This formalism consists of defining the redshifts of the two density fields in terms of a Taylor expansion around their mean and projecting them onto the celestial sphere with a Fourier analogue of the expansion parameter. This method differs from that shown in \cite{Scoccimarro:2015bla} through the incorporation of an expansion around the mean redshift of the correlator in Hyperuranion space prior to projection, rather than the incorporation of a new Fourier term in the angular correlator after projection. This allows for the direct estimation of unequal time corrections separately from those of observational effects such as RSDs.

In this paper, we derived a new method which we refer to as field level projection.  In this method, each field is assigned a reference redshift, which could be taken to be the mean redshift of the bin into which the field is sorted.  The field's individual redshifts are then displaced from this reference redshift through a Taylor expansion.  Having been so displaced, the fields are individually projected onto the celestial sphere with a Fourier analogue of their redshift displacement, before being correlated.  We showed that this formalism reveals first order corrections to the single tracer power spectrum. There are three such corrections: one accounts for deviations of the individual fields from their bin means, one accounts for the distances between their reference redshifts in the case of cross-bin analyses, and one accounts for the disparity in the time evolution of matter and biasing.

We showed that the corrections within a given redshift bin can lead to percent level corrections for fields spaced over bins with widths of the order of $100~\mathrm{Mpc~h}^{-1}$ or more.  These corrections account for deviations of the correlated fields from the mean redshift of the bin and show only a minimal dependence on the value of that redshift, instead being functions of the bin's width.  These results indicate that they may lead to a non-trivial corrections, particularly in the case of wide bins as used in some high redshift surveys.   These terms originate from allowing the two fields to be displaced independently from the bin's mean redshift, removing the assumption that their deviations from a reference point are equal and opposite, as is the case in the correlator-level projection of a power spectrum. In App.~\ref{sec:BACL}, we show that an extended version of the correlator-level projection formalism manifests these corrections after introducing a Fourier transform of deviations of the mean redshift of a power spectrum's two fields from the mean of the bin into which the power spectrum is sorted. In the appendices, we also show that these corrections can be obtained from an appropriately extended correlator-level projection for the sake of consistency checking. These extensions incorporate an additional Fourier transform which, through a change of variables, allows the fields to be displaced to different amounts from the bin's mean redshift.

We also showed that the cross-bin corrections become percent level for radial bins when the reduced ratio of their mean redshifts is of the order of $10\%$. Cross-bin correlators are usually much smaller than equal time correlators. However, we emphasize that our formalism allows for the correlation of tracers close to the border between their respective bins which would ordinarily be treated as being at their bin's mean redshifts. This substantially reduces the radial distance between the correlated fields and by extension may lead to a more significant correlation amplitude. Furthermore, integrated relativistic effects lead to cross-bin correlation that are orders of magnitude larger than those of Newtonian analyses. We are currently developing a relativistic version of our formalism which will be presented in a future paper.  Furthermore, the splitting of data into distinct redshift bins potentially reduces the information obtainable in LSS analyses.  The introduction of cross-bin correction terms provides the potential for accounting for this and reintroducing the lost information without altering the binning of a given survey.

In App.~\ref{sec:DC}, we show that both the single bin and cross-bin corrections can be obtained from an extended correlator level projection applied to an angular power spectrum.  This, together with the single bin extended correlator level projection derived in App.~\ref{sec:BACL}, constitutes a confirmation of consistency between the correlator and field level projections when they are appropriately established.

This work aimed to develop the first expression for unequal time correlators using a field level approach and highlight the introduction of first order correction terms to the single tracer power spectrum.  We will present a full analysis of field level unequal time correlations including projection effects, multi-tracer correlators and forecasts for future surveys in a future paper that is currently in preparation.

\begin{acknowledgments}
The authors thank Omar Darwish, Francesco Spezzati, and Eleonora Vanzan for helpful discussions.
AR acknowledges funding from the Italian Ministry of University and Research
(MUR) through the “Dipartimenti di eccellenza” project “Science of the Universe”. Z.V. acknowledges the support of the Kavli Foundation.
\end{acknowledgments}

\appendix

\section{Extended correlator-level Projection}
\label{sec:BACL}
We will now show that, with a suitable extension, the correlator-level projection can replicate the results of the field level projection.  This extension consists of defining a Fourier analogue of the deviation of a correlator's mean redshift from the bin's mean redshift, substituting our integrals for those used in the field level analysis, and demonstrating that this leads to equations of the same form as those shown in Section.~\ref{sec:FLP}.

We begin by recognising that the mean redshift of a given correlator can be written in terms of the mean redshift of its redshift bin as $\bar{\chi}=\bar{\chi}_{\mathrm{bin}}+\delta\bar{\chi}$, such that $\bar{\chi}$ corresponds to what was labeled $\bar{\chi}$ in Sec.~\ref{sec:CLP} and $\bar{\chi}_{\mathrm{bin}}$ corresponds to what was labeled $\bar{\chi}$ in Sec.~\ref{sec:FLP}. 
 Defining a Fourier analogue of $\delta\bar{\chi}$, which we label $p_{\hat{n}}$, and multiplying our transformed power spectrum by a factor of the bin mean redshift squared for dimensional consistency, we obtain
\begin{equation}
\begin{split}
    P(p_{\hat{n}},\mathbf{q},\bar{\chi}_{\mathrm{bin}})\equiv&\bar{\chi}_{\mathrm{bin}}^{2}\int d\delta\bar{\chi}e^{-ip_{\hat{n}}\delta\bar{\chi}}P(\mathbf{q},\bar{\chi})\\=&\bar{\chi}_{\mathrm{bin}}^{2}\int d\delta\bar{\chi}\int \frac{dk_{\hat{n}}}{2\pi}\int d\delta\chi e^{-i\delta\bar{\chi}p_{\hat{n}}}e^{i\delta\chi(k_{\hat{n}}-q_{\hat{n}})}\epsilon(\bar{\chi},\delta\chi)\mathcal{P}(k_{\hat{n}},\vec{\ell},\bar{\chi},\delta\chi)~.
    \label{eq:expandedCLClbin1}
    \end{split}
\end{equation}
Recognising that $\chi_{1}=\bar{\chi}_{\mathrm{bin}}+\delta\bar{\chi}+\frac{1}{2}\delta\chi$, such that $\delta\chi_{1}=\delta\bar{\chi}+\frac{1}{2}\delta\chi$, and $\chi_{2}=\bar{\chi}_{\mathrm{bin}}+\delta\bar{\chi}-\frac{1}{2}\delta\chi$, such that $\delta\chi_{2}=\delta\bar{\chi}-\frac{1}{2}\delta\chi$, we can substitute our integrals to obtain
\begin{equation}
\begin{split}
    P(p_{\hat{n}},\mathbf{q},\bar{\chi}_{\mathrm{bin}})=&\bar{\chi}_{\mathrm{bin}}^{2}\int \frac{dk_{\hat{n}}}{2\pi}\int d\delta\chi_{1}\int d\delta\chi_{2} e^{-i\delta\bar{\chi}p_{\hat{n}}}e^{i\delta\chi(k_{\hat{n}}-q_{\hat{n}})}\epsilon(\bar{\chi},\delta\chi)\mathcal{P}(k_{\hat{n}},\vec{\ell},\bar{\chi},\delta\chi)~.
    \label{eq:expandedCLClbin2}
    \end{split}
\end{equation}
Now, $\delta\bar{\chi}=\frac{1}{2}(\delta\chi_{1}+\delta\chi_{2})$ and $\delta\chi=\delta\chi_{1}-\delta\chi_{2}$.  Furthermore, we may recognise that $\bar{\chi}_{\mathrm{bin}}^{2}\epsilon(\chi_{1},\chi_{2})=\gamma(\chi_{1})\gamma(\chi_{2})$ in the single bin case upon Taylor expanding and remember that when expanded, $\bar{\chi}_{\mathrm{bin}}^{2}\epsilon(\chi_{1},\chi_{2})\approx 1$.  Now defining $q_{\hat{n},1}=\frac{1}{2}p_{\hat{n}}+q_{\hat{n}}$ and $q_{\hat{n},2}=\frac{1}{2}p_{\hat{n}}-q_{\hat{n}}$, Taylor expanding the power spectrum around the bin mean, and only keeping terms up to first order, we have
\begin{equation}
\begin{split}
    P(\mathbf{q}_{1},\mathbf{q}_{2},\bar{\chi}_{\mathrm{bin}})=&\int \frac{dk_{\hat{n}}}{2\pi}\int d\delta\chi_{1}\int d\delta\chi_{2} e^{-i\delta\chi_{1}(q_{\hat{n},1}+k_{\hat{n}})}e^{-i\delta\chi_{2}(q_{\hat{n},2}-k_{\hat{n}})}\left[1+F_{a}(\bar{\chi}_{\mathrm{bin}})H(\bar{\chi}_{\mathrm{bin}})\delta\bar{\chi}\right]^{2}\mathcal{P}(k_{\hat{n}},\vec{\ell},\chi_{1},\chi_{2})\\=&\int \frac{dk_{\hat{n}}}{2\pi}\left[1-iF_{a}(\bar{\chi}_{\mathrm{bin}})H(\bar{\chi}_{\mathrm{bin}})\left(\partial_{q_{\hat{n},1}}+\partial_{q_{\hat{n},2}}\right)\right]\delta^{\mathrm{D}}(q_{\hat{n},1}+k_{\hat{n}})\delta^{\mathrm{D}}(q_{\hat{n},2}-k_{\hat{n}})\mathcal{P}(k_{\hat{n}},\vec{\ell},\bar{\chi}_{\mathrm{bin}})~.
    \label{eq:expandedCLClbin2}
    \end{split}
\end{equation}
Integrating over one of the $q_{\hat{n}}$, this becomes
\begin{equation}
    P(\mathbf{q},\bar{\chi}_{\mathrm{bin}})=\left[1+iF_{a}(\bar{\chi}_{\mathrm{bin}})H(\bar{\chi}_{\mathrm{bin}})\partial_{q_{\hat{n}}}\right]\mathcal{P}(\mathbf{q},\bar{\chi}_{\mathrm{bin}})~,
\end{equation}
which is the same as the field level projection in a single bin as shown in Sec.~\ref{sec:pk}.

As with the field level projection, these corrections arise from the fact that we are allowing the two tracers to vary independently around the bin's mean redshift and generating a Fourier transform of the ensuing distribution.  In the case of the field level projection, this allowance was explicitly stated by defining separate projections for each field, while in the case of the extended correlator-level projection, it was derived by a change of variables after allowing both the fields to be displaced from the mean of their correlator and their correlator's mean to be displaced from the mean of the bin.

\section{Displacing $\mathbb{C}$}
\label{sec:DC}
The formalism in App.~\ref{sec:BACL} shows that an extended version of the correlator level projection can recreate the results of the field level projection in a single bin, but does not extend to cross-bin correlators.  We now wish to develop a formalism which directly projects the angular power spectrum while allowing the fields to be located in separate bins.  

We begin by applying the operation defined in Eq.~\eqref{eq:deltac} to the function defined in Eq.~\eqref{eq:UTCl1} and dividing by a factor of $\bar{\chi}_{1}\bar{\chi}_{2}$ for dimensionality:
\begin{equation}
\begin{split}
P(\mathbf{q}_{1},\mathbf{q}_{2},\bar{\chi}_{1},\bar{\chi}_{2})\equiv& \bar{\chi}_{1}\bar{\chi}_{2}\int d \delta\chi_{1} e^{-i\delta\chi_{1} q_{\hat{n},1}}\int d \delta\chi_{2} e^{-i\delta\chi_{2} q_{\hat{n},2}}\mathbb{C}(\vec{\ell},\chi_{1},\chi_{2})~,\\=&(2\pi)^{3}\int\frac{dk_{\hat{n}}}{2\pi}\int d\delta\chi_{1}\int d\delta\chi_{2}\upsilon(\bar{\chi},\delta\chi)e^{ik_{\hat{n}}(\bar{\chi}_{1}-\bar{\chi}_{2})}e^{i\delta\chi_{1} (k_{\hat{n}}-q_{\hat{n},1})}e^{-i\delta\chi_{2} (k_{\hat{n}}+q_{\hat{n},2})}\mathcal{P}(k_{\hat{n}},\vec{\ell},\chi_{1},\chi_{2})~,
\label{eq:CLCl2}
\end{split}
\end{equation}
where $\upsilon(\chi_{1},\chi_{2})=\sqrt{\epsilon(\chi_{1},\chi_{2})}$.  Introducing our Taylor expansion to Eq.~\eqref{eq:CLCl2} and noting that, as was the case with $\gamma$ in Sec.~\ref{sec:PontoCS}, $\upsilon(\bar{\chi}_{1},\bar{\chi}_{2})=1$ and the higher order terms in its expansion vanish, we have
\begin{equation}
\begin{split}
P(\mathbf{q}_{1},\mathbf{q}_{2},\bar{\chi}_{1},\bar{\chi}_{2})=&(2\pi)^{3}\int\frac{dk_{\hat{n}}}{2\pi}e^{ik_{\hat{n}}(\bar{\chi}_{1}-\bar{\chi}_{2})}\int d\delta\chi_{1}\int d\delta\chi_{2}\upsilon(\bar{\chi},\delta\chi)e^{i\delta\chi_{1} (k_{\hat{n}}-q_{\hat{n},1})}e^{-i\delta\chi_{2} (k_{\hat{n}}+q_{\hat{n},2})}\\&\times\left[1+F_{a}(\bar{\chi}_{1})H(\bar{\chi}_{1})\delta\chi_{1}\right]\left[1+F_{a}(\bar{\chi}_{2})H(\bar{\chi}_{2})\delta\chi_{2}\right]\mathcal{P}(k_{\hat{n}},\vec{\ell},\bar{\chi}_{1},\bar{\chi}_{2})\\=&(2\pi)^{3}\int\frac{dk_{\hat{n}}}{2\pi}e^{ik_{\hat{n}}(\bar{\chi}_{1}-\bar{\chi}_{2})}\left[1+iF_{a}(\bar{\chi}_{1})H(\bar{\chi}_{1})\partial_{q_{\hat{n},1}}+iF_{a}(\bar{\chi}_{2})H(\bar{\chi}_{2})\partial_{q_{\hat{n},2}}\right]\\&\times\delta^{\mathrm{D}}(k_{\hat{n}}-q_{\hat{n},1})\delta^{\mathrm{D}}(k_{\hat{n}}+q_{\hat{n},2})\mathcal{P}(k_{\hat{n}},\vec{\ell},\bar{\chi}_{1},\bar{\chi}_{2})~.
\label{eq:CLCl2}
\end{split}
\end{equation}
Integrating over one of the momenta then gives the same results as were obtained in Sec.~\ref{sec:pk} and setting $\bar{\chi}_{1}=\bar{\chi}_{2}$ gives the same results as were obtained in App.~\ref{sec:BACL}, demonstrating that the formalism of operating directly upon the angular power spectrum obtains the same results as the other methods discussed in this paper.

\bibliography{bibliography}

\begin{thebibliography}{71}%
\makeatletter
\providecommand \@ifxundefined [1]{%
 \@ifx{#1\undefined}
}%
\providecommand \@ifnum [1]{%
 \ifnum #1\expandafter \@firstoftwo
 \else \expandafter \@secondoftwo
 \fi
}%
\providecommand \@ifx [1]{%
 \ifx #1\expandafter \@firstoftwo
 \else \expandafter \@secondoftwo
 \fi
}%
\providecommand \natexlab [1]{#1}%
\providecommand \enquote  [1]{``#1''}%
\providecommand \bibnamefont  [1]{#1}%
\providecommand \bibfnamefont [1]{#1}%
\providecommand \citenamefont [1]{#1}%
\providecommand \href@noop [0]{\@secondoftwo}%
\providecommand \href [0]{\begingroup \@sanitize@url \@href}%
\providecommand \@href[1]{\@@startlink{#1}\@@href}%
\providecommand \@@href[1]{\endgroup#1\@@endlink}%
\providecommand \@sanitize@url [0]{\catcode `\\12\catcode `\$12\catcode
  `\&12\catcode `\#12\catcode `\^12\catcode `\_12\catcode `\%12\relax}%
\providecommand \@@startlink[1]{}%
\providecommand \@@endlink[0]{}%
\providecommand \url  [0]{\begingroup\@sanitize@url \@url }%
\providecommand \@url [1]{\endgroup\@href {#1}{\urlprefix }}%
\providecommand \urlprefix  [0]{URL }%
\providecommand \Eprint [0]{\href }%
\providecommand \doibase [0]{https://doi.org/}%
\providecommand \selectlanguage [0]{\@gobble}%
\providecommand \bibinfo  [0]{\@secondoftwo}%
\providecommand \bibfield  [0]{\@secondoftwo}%
\providecommand \translation [1]{[#1]}%
\providecommand \BibitemOpen [0]{}%
\providecommand \bibitemStop [0]{}%
\providecommand \bibitemNoStop [0]{.\EOS\space}%
\providecommand \EOS [0]{\spacefactor3000\relax}%
\providecommand \BibitemShut  [1]{\csname bibitem#1\endcsname}%
\let\auto@bib@innerbib\@empty
\bibitem [{\citenamefont {{Laureijs}}\ \emph {et~al.}(2011)\citenamefont
  {{Laureijs}},  \emph {et~al.}}]{2011arXiv1110.3193L}%
  \BibitemOpen
  \bibfield  {author} {\bibinfo {author} {\bibfnamefont {R.}~\bibnamefont
  {{Laureijs}}}, , \emph {et~al.},\ }\bibfield  {title} {\bibinfo {title}
  {{Euclid Definition Study Report}},\ }\href@noop {} {\bibfield  {journal}
  {\bibinfo  {journal} {arXiv e-prints}\ ,\ \bibinfo {eid} {arXiv:1110.3193}}
  (\bibinfo {year} {2011})},\ \Eprint {https://arxiv.org/abs/1110.3193}
  {arXiv:1110.3193 [astro-ph.CO]} \BibitemShut {NoStop}%
\bibitem [{\citenamefont {Dor\'e}\ \emph {et~al.}(2014)\citenamefont {Dor\'e}
  \emph {et~al.}}]{Dore:2014cca}%
  \BibitemOpen
  \bibfield  {author} {\bibinfo {author} {\bibfnamefont {O.}~\bibnamefont
  {Dor\'e}} \emph {et~al.},\ }\bibfield  {title} {\bibinfo {title} {{Cosmology
  with the SPHEREX All-Sky Spectral Survey}},\ }\href@noop {} {\  (\bibinfo
  {year} {2014})},\ \Eprint {https://arxiv.org/abs/1412.4872} {arXiv:1412.4872
  [astro-ph.CO]} \BibitemShut {NoStop}%
\bibitem [{\citenamefont {{Ivezi{\'c}}}\ \emph {et~al.}(2019)\citenamefont
  {{Ivezi{\'c}}} \emph {et~al.}}]{2019ApJ...873..111I}%
  \BibitemOpen
  \bibfield  {author} {\bibinfo {author} {\bibfnamefont {{\v{Z}}.}~\bibnamefont
  {{Ivezi{\'c}}}} \emph {et~al.},\ }\bibfield  {title} {\bibinfo {title}
  {{LSST: From Science Drivers to Reference Design and Anticipated Data
  Products}},\ }\href {https://doi.org/10.3847/1538-4357/ab042c} {\bibfield
  {journal} {\bibinfo  {journal} {\apj}\ }\textbf {\bibinfo {volume} {873}},\
  \bibinfo {eid} {111} (\bibinfo {year} {2019})},\ \Eprint
  {https://arxiv.org/abs/0805.2366} {arXiv:0805.2366 [astro-ph]} \BibitemShut
  {NoStop}%
\bibitem [{\citenamefont {Hamana}\ \emph {et~al.}(2020)\citenamefont {Hamana}
  \emph {et~al.}}]{Hamana:2019etx}%
  \BibitemOpen
  \bibfield  {author} {\bibinfo {author} {\bibfnamefont {T.}~\bibnamefont
  {Hamana}} \emph {et~al.},\ }\bibfield  {title} {\bibinfo {title}
  {{Cosmological constraints from cosmic shear two-point correlation functions
  with HSC survey first-year data}},\ }\href
  {https://doi.org/10.1093/pasj/psz138} {\bibfield  {journal} {\bibinfo
  {journal} {Publ. Astron. Soc. Jap.}\ }\textbf {\bibinfo {volume} {72}},\
  \bibinfo {pages} {Publications of the Astronomical Society of Japan, Volume
  72, Issue 1, February 2020, 16, https://doi.org/10.1093/pasj/psz138}
  (\bibinfo {year} {2020})},\ \Eprint {https://arxiv.org/abs/1906.06041}
  {arXiv:1906.06041 [astro-ph.CO]} \BibitemShut {NoStop}%
\bibitem [{\citenamefont {Yahya}\ \emph {et~al.}(2015)\citenamefont {Yahya}
  \emph {et~al.}}]{10.1093/mnras/stv695}%
  \BibitemOpen
  \bibfield  {author} {\bibinfo {author} {\bibfnamefont {S.}~\bibnamefont
  {Yahya}} \emph {et~al.},\ }\bibfield  {title} {\bibinfo {title}
  {{Cosmological performance of SKA Hi galaxy surveys}},\ }\href
  {https://doi.org/10.1093/mnras/stv695} {\bibfield  {journal} {\bibinfo
  {journal} {Monthly Notices of the Royal Astronomical Society}\ }\textbf
  {\bibinfo {volume} {450}},\ \bibinfo {pages} {2251} (\bibinfo {year}
  {2015})},\ \Eprint
  {https://arxiv.org/abs/https://academic.oup.com/mnras/article-pdf/450/3/2251/18504481/stv695.pdf}
  {https://academic.oup.com/mnras/article-pdf/450/3/2251/18504481/stv695.pdf}
  \BibitemShut {NoStop}%
\bibitem [{\citenamefont {Wang}\ \emph {et~al.}(2021)\citenamefont {Wang} \emph
  {et~al.}}]{Wang:2021moa}%
  \BibitemOpen
  \bibfield  {author} {\bibinfo {author} {\bibfnamefont {Z.}~\bibnamefont
  {Wang}} \emph {et~al.},\ }\bibfield  {title} {\bibinfo {title} {{The
  clustering of galaxies in the DESI imaging legacy surveys DR8: I. The
  luminosity and color dependent intrinsic clustering}},\ }\href
  {https://doi.org/10.1007/s11433-021-1707-6} {\bibfield  {journal} {\bibinfo
  {journal} {Sci. China Phys. Mech. Astron.}\ }\textbf {\bibinfo {volume}
  {64}},\ \bibinfo {pages} {289811} (\bibinfo {year} {2021})},\ \Eprint
  {https://arxiv.org/abs/2106.14159} {arXiv:2106.14159 [astro-ph.GA]}
  \BibitemShut {NoStop}%
\bibitem [{\citenamefont {Pullen}\ \emph {et~al.}(2015)\citenamefont {Pullen},
  \citenamefont {Hirata}, \citenamefont {Doré},\ and\ \citenamefont
  {Raccanelli}}]{interloperbias}%
  \BibitemOpen
  \bibfield  {author} {\bibinfo {author} {\bibfnamefont {A.}~\bibnamefont
  {Pullen}}, \bibinfo {author} {\bibfnamefont {C.}~\bibnamefont {Hirata}},
  \bibinfo {author} {\bibfnamefont {O.}~\bibnamefont {Doré}},\ and\ \bibinfo
  {author} {\bibfnamefont {A.}~\bibnamefont {Raccanelli}},\ }\bibfield  {title}
  {\bibinfo {title} {Interloper bias in future large-scale structure surveys},\
  }\href {https://doi.org/10.1093/pasj/psv118} {\bibfield  {journal} {\bibinfo
  {journal} {Publications of the Astronomical Society of Japan}\ }\textbf
  {\bibinfo {volume} {68}} (\bibinfo {year} {2015})}\BibitemShut {NoStop}%
\bibitem [{\citenamefont {Richard}\ \emph {et~al.}(2019)\citenamefont {Richard}
  \emph {et~al.}}]{4most}%
  \BibitemOpen
  \bibfield  {author} {\bibinfo {author} {\bibfnamefont {J.}~\bibnamefont
  {Richard}} \emph {et~al.},\ }\href {https://doi.org/10.18727/0722-6691/5127}
  {\bibinfo {title} {4most consortium survey 8: Cosmology redshift survey
  (crs)}} (\bibinfo {year} {2019})\BibitemShut {NoStop}%
\bibitem [{\citenamefont {Takada}\ \emph {et~al.}(2014)\citenamefont {Takada}
  \emph {et~al.}}]{10.1093/pasj/pst019}%
  \BibitemOpen
  \bibfield  {author} {\bibinfo {author} {\bibfnamefont {M.}~\bibnamefont
  {Takada}} \emph {et~al.},\ }\bibfield  {title} {\bibinfo {title}
  {{Extragalactic science, cosmology, and Galactic archaeology with the Subaru
  Prime Focus Spectrograph}},\ }\bibfield  {journal} {\bibinfo  {journal}
  {Publications of the Astronomical Society of Japan}\ }\textbf {\bibinfo
  {volume} {66}},\ \href {https://doi.org/10.1093/pasj/pst019}
  {10.1093/pasj/pst019} (\bibinfo {year} {2014}),\ \bibinfo {note} {r1},\
  \Eprint
  {https://arxiv.org/abs/https://academic.oup.com/pasj/article-pdf/66/1/R1/4425324/pst019.pdf}
  {https://academic.oup.com/pasj/article-pdf/66/1/R1/4425324/pst019.pdf}
  \BibitemShut {NoStop}%
\bibitem [{\citenamefont {Hildebrandt}\ \emph {et~al.}(2021)\citenamefont
  {Hildebrandt} \emph {et~al.}}]{Hildebrandt:2020rno}%
  \BibitemOpen
  \bibfield  {author} {\bibinfo {author} {\bibfnamefont {H.}~\bibnamefont
  {Hildebrandt}} \emph {et~al.},\ }\bibfield  {title} {\bibinfo {title}
  {{KiDS-1000 catalogue: Redshift distributions and their calibration}},\
  }\href {https://doi.org/10.1051/0004-6361/202039018} {\bibfield  {journal}
  {\bibinfo  {journal} {Astron. Astrophys.}\ }\textbf {\bibinfo {volume}
  {647}},\ \bibinfo {pages} {A124} (\bibinfo {year} {2021})},\ \Eprint
  {https://arxiv.org/abs/2007.15635} {arXiv:2007.15635 [astro-ph.CO]}
  \BibitemShut {NoStop}%
\bibitem [{\citenamefont {{Dawson}}\ \emph {et~al.}()\citenamefont {{Dawson}}
  \emph {et~al.}}]{2013AJ....145...10D}%
  \BibitemOpen
  \bibfield  {author} {\bibinfo {author} {\bibfnamefont {K.~S.}\ \bibnamefont
  {{Dawson}}} \emph {et~al.},\ }\bibfield  {title} {\bibinfo {title} {{The
  Baryon Oscillation Spectroscopic Survey of SDSS-III}},\ }\bibfield  {journal}
  {\bibinfo  {journal} {The Astronomical Journal}\ }\textbf {\bibinfo {volume}
  {145}},\ \href {https://doi.org/10.1088/0004-6256/145/1/10}
  {10.1088/0004-6256/145/1/10},\ \Eprint {https://arxiv.org/abs/1208.0022}
  {arXiv:1208.0022 [astro-ph.CO]} \BibitemShut {NoStop}%
\bibitem [{\citenamefont {Icaza-Lizaola}\ \emph {et~al.}(2019)\citenamefont
  {Icaza-Lizaola} \emph {et~al.}}]{10.1093/mnras/stz3602}%
  \BibitemOpen
  \bibfield  {author} {\bibinfo {author} {\bibfnamefont {M.}~\bibnamefont
  {Icaza-Lizaola}} \emph {et~al.},\ }\bibfield  {title} {\bibinfo {title} {{The
  clustering of the SDSS-IV extended Baryon Oscillation Spectroscopic Survey
  DR14 LRG sample: structure growth rate measurement from the anisotropic LRG
  correlation function in the redshift range 0.6<z< 1.0}},\ }\href
  {https://doi.org/10.1093/mnras/stz3602} {\bibfield  {journal} {\bibinfo
  {journal} {Monthly Notices of the Royal Astronomical Society}\ }\textbf
  {\bibinfo {volume} {492}},\ \bibinfo {pages} {4189} (\bibinfo {year}
  {2019})},\ \Eprint
  {https://arxiv.org/abs/https://academic.oup.com/mnras/article-pdf/492/3/4189/32217846/stz3602.pdf}
  {https://academic.oup.com/mnras/article-pdf/492/3/4189/32217846/stz3602.pdf}
  \BibitemShut {NoStop}%
\bibitem [{\citenamefont {{Weinberg}}\ \emph {et~al.}(2013)\citenamefont
  {{Weinberg}} \emph {et~al.}}]{2013PhR...530...87W}%
  \BibitemOpen
  \bibfield  {author} {\bibinfo {author} {\bibfnamefont {D.~H.}\ \bibnamefont
  {{Weinberg}}} \emph {et~al.},\ }\bibfield  {title} {\bibinfo {title}
  {{Observational probes of cosmic acceleration}},\ }\href
  {https://doi.org/10.1016/j.physrep.2013.05.001} {\bibfield  {journal}
  {\bibinfo  {journal} {Physics Report}\ }\textbf {\bibinfo {volume} {530}},\
  \bibinfo {pages} {87} (\bibinfo {year} {2013})},\ \Eprint
  {https://arxiv.org/abs/1201.2434} {arXiv:1201.2434 [astro-ph.CO]}
  \BibitemShut {NoStop}%
\bibitem [{\citenamefont {Martinelli}\ \emph {et~al.}(2021)\citenamefont
  {Martinelli} \emph {et~al.}}]{Euclid:2021cfn}%
  \BibitemOpen
  \bibfield  {author} {\bibinfo {author} {\bibfnamefont {M.}~\bibnamefont
  {Martinelli}} \emph {et~al.} (\bibinfo {collaboration} {Euclid}),\ }\bibfield
   {title} {\bibinfo {title} {{Euclid: Constraining dark energy coupled to
  electromagnetism using astrophysical and laboratory data}},\ }\href
  {https://doi.org/10.1051/0004-6361/202141353} {\bibfield  {journal} {\bibinfo
   {journal} {Astron. Astrophys.}\ }\textbf {\bibinfo {volume} {654}},\
  \bibinfo {pages} {A148} (\bibinfo {year} {2021})},\ \Eprint
  {https://arxiv.org/abs/2105.09746} {arXiv:2105.09746 [astro-ph.CO]}
  \BibitemShut {NoStop}%
\bibitem [{\citenamefont {Piga}\ \emph {et~al.}(2023)\citenamefont {Piga},
  \citenamefont {Marinucci}, \citenamefont {D'Amico}, \citenamefont {Pietroni},
  \citenamefont {Vernizzi},\ and\ \citenamefont {Wright}}]{Piga:2022mge}%
  \BibitemOpen
  \bibfield  {author} {\bibinfo {author} {\bibfnamefont {L.}~\bibnamefont
  {Piga}}, \bibinfo {author} {\bibfnamefont {M.}~\bibnamefont {Marinucci}},
  \bibinfo {author} {\bibfnamefont {G.}~\bibnamefont {D'Amico}}, \bibinfo
  {author} {\bibfnamefont {M.}~\bibnamefont {Pietroni}}, \bibinfo {author}
  {\bibfnamefont {F.}~\bibnamefont {Vernizzi}},\ and\ \bibinfo {author}
  {\bibfnamefont {B.~S.}\ \bibnamefont {Wright}},\ }\bibfield  {title}
  {\bibinfo {title} {{Constraints on modified gravity from the BOSS galaxy
  survey}},\ }\href {https://doi.org/10.1088/1475-7516/2023/04/038} {\bibfield
  {journal} {\bibinfo  {journal} {JCAP}\ }\textbf {\bibinfo {volume} {04}},\
  \bibinfo {pages} {038}},\ \Eprint {https://arxiv.org/abs/2211.12523}
  {arXiv:2211.12523 [astro-ph.CO]} \BibitemShut {NoStop}%
\bibitem [{\citenamefont {{Holsclaw}}\ \emph {et~al.}(2011)\citenamefont
  {{Holsclaw}} \emph {et~al.}}]{2011PhRvD..84h3501H}%
  \BibitemOpen
  \bibfield  {author} {\bibinfo {author} {\bibfnamefont {T.}~\bibnamefont
  {{Holsclaw}}} \emph {et~al.},\ }\bibfield  {title} {\bibinfo {title}
  {{Nonparametric reconstruction of the dark energy equation of state from
  diverse data sets}},\ }\href {https://doi.org/10.1103/PhysRevD.84.083501}
  {\bibfield  {journal} {\bibinfo  {journal} {\prd}\ }\textbf {\bibinfo
  {volume} {84}},\ \bibinfo {eid} {083501} (\bibinfo {year} {2011})},\ \Eprint
  {https://arxiv.org/abs/1104.2041} {arXiv:1104.2041 [astro-ph.CO]}
  \BibitemShut {NoStop}%
\bibitem [{\citenamefont {{Giannantonio}}\ \emph {et~al.}(2012)\citenamefont
  {{Giannantonio}} \emph {et~al.}}]{2012MNRAS.422.2854G}%
  \BibitemOpen
  \bibfield  {author} {\bibinfo {author} {\bibfnamefont {T.}~\bibnamefont
  {{Giannantonio}}} \emph {et~al.},\ }\bibfield  {title} {\bibinfo {title}
  {{Constraining primordial non-Gaussianity with future galaxy surveys}},\
  }\href {https://doi.org/10.1111/j.1365-2966.2012.20604.x} {\bibfield
  {journal} {\bibinfo  {journal} {Mon. Not. Roy. Astron. Soc.}\ }\textbf
  {\bibinfo {volume} {422}},\ \bibinfo {pages} {2854} (\bibinfo {year}
  {2012})},\ \Eprint {https://arxiv.org/abs/1109.0958} {arXiv:1109.0958
  [astro-ph.CO]} \BibitemShut {NoStop}%
\bibitem [{\citenamefont {Karagiannis}\ \emph {et~al.}(2018)\citenamefont
  {Karagiannis} \emph {et~al.}}]{Karagiannis:2018jdt}%
  \BibitemOpen
  \bibfield  {author} {\bibinfo {author} {\bibfnamefont {D.}~\bibnamefont
  {Karagiannis}} \emph {et~al.},\ }\bibfield  {title} {\bibinfo {title}
  {{Constraining primordial non-Gaussianity with bispectrum and power spectrum
  from upcoming optical and radio surveys}},\ }\href
  {https://doi.org/10.1093/mnras/sty1029} {\bibfield  {journal} {\bibinfo
  {journal} {Mon. Not. Roy. Astron. Soc.}\ }\textbf {\bibinfo {volume} {478}},\
  \bibinfo {pages} {1341} (\bibinfo {year} {2018})},\ \Eprint
  {https://arxiv.org/abs/1801.09280} {arXiv:1801.09280 [astro-ph.CO]}
  \BibitemShut {NoStop}%
\bibitem [{\citenamefont {Yamauchi}\ \emph {et~al.}(2014)\citenamefont
  {Yamauchi}, \citenamefont {Takahashi},\ and\ \citenamefont
  {Oguri}}]{Yamauchi:2014ioa}%
  \BibitemOpen
  \bibfield  {author} {\bibinfo {author} {\bibfnamefont {D.}~\bibnamefont
  {Yamauchi}}, \bibinfo {author} {\bibfnamefont {K.}~\bibnamefont
  {Takahashi}},\ and\ \bibinfo {author} {\bibfnamefont {M.}~\bibnamefont
  {Oguri}},\ }\bibfield  {title} {\bibinfo {title} {{Constraining primordial
  non-Gaussianity via a multitracer technique with surveys by Euclid and the
  Square Kilometre Array}},\ }\href
  {https://doi.org/10.1103/PhysRevD.90.083520} {\bibfield  {journal} {\bibinfo
  {journal} {Phys. Rev. D}\ }\textbf {\bibinfo {volume} {90}},\ \bibinfo
  {pages} {083520} (\bibinfo {year} {2014})},\ \Eprint
  {https://arxiv.org/abs/1407.5453} {arXiv:1407.5453 [astro-ph.CO]}
  \BibitemShut {NoStop}%
\bibitem [{\citenamefont {Brown}\ \emph {et~al.}(2024)\citenamefont {Brown}
  \emph {et~al.}}]{Brown:2024dmv}%
  \BibitemOpen
  \bibfield  {author} {\bibinfo {author} {\bibfnamefont {Z.}~\bibnamefont
  {Brown}} \emph {et~al.},\ }\bibfield  {title} {\bibinfo {title}
  {{Constraining primordial non-Gaussianity from the large scale structure
  two-point and three-point correlation functions}},\ }\href@noop {} {\
  (\bibinfo {year} {2024})},\ \Eprint {https://arxiv.org/abs/2403.18789}
  {arXiv:2403.18789 [astro-ph.CO]} \BibitemShut {NoStop}%
\bibitem [{\citenamefont {Hahn}\ \emph {et~al.}(2020)\citenamefont {Hahn},
  \citenamefont {Villaescusa-Navarro}, \citenamefont {Castorina},\ and\
  \citenamefont {Scoccimarro}}]{Hahn:2019zob}%
  \BibitemOpen
  \bibfield  {author} {\bibinfo {author} {\bibfnamefont {C.}~\bibnamefont
  {Hahn}}, \bibinfo {author} {\bibfnamefont {F.}~\bibnamefont
  {Villaescusa-Navarro}}, \bibinfo {author} {\bibfnamefont {E.}~\bibnamefont
  {Castorina}},\ and\ \bibinfo {author} {\bibfnamefont {R.}~\bibnamefont
  {Scoccimarro}},\ }\bibfield  {title} {\bibinfo {title} {{Constraining $M_\nu$
  with the bispectrum. Part I. Breaking parameter degeneracies}},\ }\href
  {https://doi.org/10.1088/1475-7516/2020/03/040} {\bibfield  {journal}
  {\bibinfo  {journal} {JCAP}\ }\textbf {\bibinfo {volume} {03}},\ \bibinfo
  {pages} {040}},\ \Eprint {https://arxiv.org/abs/1909.11107} {arXiv:1909.11107
  [astro-ph.CO]} \BibitemShut {NoStop}%
\bibitem [{\citenamefont {Cerbolini}\ \emph {et~al.}(2013)\citenamefont
  {Cerbolini} \emph {et~al.}}]{Cerbolini:2013uya}%
  \BibitemOpen
  \bibfield  {author} {\bibinfo {author} {\bibfnamefont {M.~C.~A.}\
  \bibnamefont {Cerbolini}} \emph {et~al.},\ }\bibfield  {title} {\bibinfo
  {title} {{Constraining neutrino properties with a Euclid-like galaxy cluster
  survey}},\ }\href {https://doi.org/10.1088/1475-7516/2013/06/020} {\bibfield
  {journal} {\bibinfo  {journal} {JCAP}\ }\textbf {\bibinfo {volume} {06}},\
  \bibinfo {pages} {020}},\ \Eprint {https://arxiv.org/abs/1303.4550}
  {arXiv:1303.4550 [astro-ph.CO]} \BibitemShut {NoStop}%
\bibitem [{\citenamefont {Adamek}\ \emph {et~al.}(2023)\citenamefont {Adamek}
  \emph {et~al.}}]{Euclid:2022qde}%
  \BibitemOpen
  \bibfield  {author} {\bibinfo {author} {\bibfnamefont {J.}~\bibnamefont
  {Adamek}} \emph {et~al.} (\bibinfo {collaboration} {Euclid}),\ }\bibfield
  {title} {\bibinfo {title} {{Euclid: Modelling massive neutrinos in cosmology
  \textendash{} a code comparison}},\ }\href
  {https://doi.org/10.1088/1475-7516/2023/06/035} {\bibfield  {journal}
  {\bibinfo  {journal} {JCAP}\ }\textbf {\bibinfo {volume} {06}},\ \bibinfo
  {pages} {035}},\ \Eprint {https://arxiv.org/abs/2211.12457} {arXiv:2211.12457
  [astro-ph.CO]} \BibitemShut {NoStop}%
\bibitem [{\citenamefont {Dvorkin}\ \emph {et~al.}(2019)\citenamefont {Dvorkin}
  \emph {et~al.}}]{Dvorkin:2019jgs}%
  \BibitemOpen
  \bibfield  {author} {\bibinfo {author} {\bibfnamefont {C.}~\bibnamefont
  {Dvorkin}} \emph {et~al.},\ }\bibfield  {title} {\bibinfo {title} {{Neutrino
  Mass from Cosmology: Probing Physics Beyond the Standard Model}},\
  }\href@noop {} {\  (\bibinfo {year} {2019})},\ \Eprint
  {https://arxiv.org/abs/1903.03689} {arXiv:1903.03689 [astro-ph.CO]}
  \BibitemShut {NoStop}%
\bibitem [{\citenamefont {Nesseris}\ \emph {et~al.}(2022)\citenamefont
  {Nesseris} \emph {et~al.}}]{Euclid:2021frk}%
  \BibitemOpen
  \bibfield  {author} {\bibinfo {author} {\bibfnamefont {S.}~\bibnamefont
  {Nesseris}} \emph {et~al.} (\bibinfo {collaboration} {Euclid}),\ }\bibfield
  {title} {\bibinfo {title} {{Euclid: Forecast constraints on consistency tests
  of the \ensuremath{\Lambda}CDM model}},\ }\href
  {https://doi.org/10.1051/0004-6361/202142503} {\bibfield  {journal} {\bibinfo
   {journal} {Astron. Astrophys.}\ }\textbf {\bibinfo {volume} {660}},\
  \bibinfo {pages} {A67} (\bibinfo {year} {2022})},\ \Eprint
  {https://arxiv.org/abs/2110.11421} {arXiv:2110.11421 [astro-ph.CO]}
  \BibitemShut {NoStop}%
\bibitem [{\citenamefont {{Camarena}}\ \emph {et~al.}(2023)\citenamefont
  {{Camarena}} \emph {et~al.}}]{2023A&A...671A..68C}%
  \BibitemOpen
  \bibfield  {author} {\bibinfo {author} {\bibfnamefont {D.}~\bibnamefont
  {{Camarena}}} \emph {et~al.},\ }\bibfield  {title} {\bibinfo {title}
  {{Euclid: Testing the Copernican principle with next-generation surveys}},\
  }\href {https://doi.org/10.1051/0004-6361/202244557} {\bibfield  {journal}
  {\bibinfo  {journal} {Astronomy and Astrophysics}\ }\textbf {\bibinfo
  {volume} {671}},\ \bibinfo {eid} {A68} (\bibinfo {year} {2023})},\ \Eprint
  {https://arxiv.org/abs/2207.09995} {arXiv:2207.09995 [astro-ph.CO]}
  \BibitemShut {NoStop}%
\bibitem [{\citenamefont {Pandey}\ and\ \citenamefont
  {Sarkar}(2021)}]{Pandey:2021qqp}%
  \BibitemOpen
  \bibfield  {author} {\bibinfo {author} {\bibfnamefont {B.}~\bibnamefont
  {Pandey}}\ and\ \bibinfo {author} {\bibfnamefont {S.}~\bibnamefont
  {Sarkar}},\ }\bibfield  {title} {\bibinfo {title} {{Testing homogeneity of
  the galaxy distribution in the SDSS using Renyi entropy}}\ }\href
  {https://doi.org/10.1088/1475-7516/2021/07/019}
  {10.1088/1475-7516/2021/07/019} (\bibinfo {year} {2021}),\ \Eprint
  {https://arxiv.org/abs/2103.11954} {arXiv:2103.11954 [astro-ph.CO]}
  \BibitemShut {NoStop}%
\bibitem [{\citenamefont {Zunckel}\ \emph {et~al.}(2011)\citenamefont
  {Zunckel}, \citenamefont {Huterer},\ and\ \citenamefont
  {Starkman}}]{Zunckel:2010yq}%
  \BibitemOpen
  \bibfield  {author} {\bibinfo {author} {\bibfnamefont {C.}~\bibnamefont
  {Zunckel}}, \bibinfo {author} {\bibfnamefont {D.}~\bibnamefont {Huterer}},\
  and\ \bibinfo {author} {\bibfnamefont {G.~D.}\ \bibnamefont {Starkman}},\
  }\bibfield  {title} {\bibinfo {title} {{Testing the statistical isotropy of
  large scale structure with multipole vectors}},\ }\href
  {https://doi.org/10.1103/PhysRevD.84.043005} {\bibfield  {journal} {\bibinfo
  {journal} {Phys. Rev. D}\ }\textbf {\bibinfo {volume} {84}},\ \bibinfo
  {pages} {043005} (\bibinfo {year} {2011})},\ \Eprint
  {https://arxiv.org/abs/1009.4701} {arXiv:1009.4701 [astro-ph.CO]}
  \BibitemShut {NoStop}%
\bibitem [{\citenamefont {Bernardeau}\ \emph {et~al.}(2002)\citenamefont
  {Bernardeau}, \citenamefont {Colombi}, \citenamefont {Gaztanaga},\ and\
  \citenamefont {Scoccimarro}}]{Bernardeau:2001qr}%
  \BibitemOpen
  \bibfield  {author} {\bibinfo {author} {\bibfnamefont {F.}~\bibnamefont
  {Bernardeau}}, \bibinfo {author} {\bibfnamefont {S.}~\bibnamefont {Colombi}},
  \bibinfo {author} {\bibfnamefont {E.}~\bibnamefont {Gaztanaga}},\ and\
  \bibinfo {author} {\bibfnamefont {R.}~\bibnamefont {Scoccimarro}},\
  }\bibfield  {title} {\bibinfo {title} {{Large scale structure of the universe
  and cosmological perturbation theory}},\ }\href
  {https://doi.org/10.1016/S0370-1573(02)00135-7} {\bibfield  {journal}
  {\bibinfo  {journal} {Phys. Rept.}\ }\textbf {\bibinfo {volume} {367}},\
  \bibinfo {pages} {1} (\bibinfo {year} {2002})},\ \Eprint
  {https://arxiv.org/abs/astro-ph/0112551} {arXiv:astro-ph/0112551}
  \BibitemShut {NoStop}%
\bibitem [{\citenamefont {Blas}\ \emph
  {et~al.}(2016{\natexlab{a}})\citenamefont {Blas}, \citenamefont {Garny},
  \citenamefont {Ivanov},\ and\ \citenamefont {Sibiryakov}}]{Blas:2015qsi}%
  \BibitemOpen
  \bibfield  {author} {\bibinfo {author} {\bibfnamefont {D.}~\bibnamefont
  {Blas}}, \bibinfo {author} {\bibfnamefont {M.}~\bibnamefont {Garny}},
  \bibinfo {author} {\bibfnamefont {M.~M.}\ \bibnamefont {Ivanov}},\ and\
  \bibinfo {author} {\bibfnamefont {S.}~\bibnamefont {Sibiryakov}},\ }\bibfield
   {title} {\bibinfo {title} {{Time-Sliced Perturbation Theory for Large Scale
  Structure I: General Formalism}},\ }\href
  {https://doi.org/10.1088/1475-7516/2016/07/052} {\bibfield  {journal}
  {\bibinfo  {journal} {JCAP}\ }\textbf {\bibinfo {volume} {07}},\ \bibinfo
  {pages} {052}},\ \Eprint {https://arxiv.org/abs/1512.05807} {arXiv:1512.05807
  [astro-ph.CO]} \BibitemShut {NoStop}%
\bibitem [{\citenamefont {Blas}\ \emph
  {et~al.}(2016{\natexlab{b}})\citenamefont {Blas}, \citenamefont {Garny},
  \citenamefont {Ivanov},\ and\ \citenamefont {Sibiryakov}}]{Blas:2016sfa}%
  \BibitemOpen
  \bibfield  {author} {\bibinfo {author} {\bibfnamefont {D.}~\bibnamefont
  {Blas}}, \bibinfo {author} {\bibfnamefont {M.}~\bibnamefont {Garny}},
  \bibinfo {author} {\bibfnamefont {M.~M.}\ \bibnamefont {Ivanov}},\ and\
  \bibinfo {author} {\bibfnamefont {S.}~\bibnamefont {Sibiryakov}},\ }\bibfield
   {title} {\bibinfo {title} {{Time-Sliced Perturbation Theory II: Baryon
  Acoustic Oscillations and Infrared Resummation}},\ }\href
  {https://doi.org/10.1088/1475-7516/2016/07/028} {\bibfield  {journal}
  {\bibinfo  {journal} {JCAP}\ }\textbf {\bibinfo {volume} {07}},\ \bibinfo
  {pages} {028}},\ \Eprint {https://arxiv.org/abs/1605.02149} {arXiv:1605.02149
  [astro-ph.CO]} \BibitemShut {NoStop}%
\bibitem [{\citenamefont {Blas}\ \emph {et~al.}(2015)\citenamefont {Blas} \emph
  {et~al.}}]{Blas:2015tla}%
  \BibitemOpen
  \bibfield  {author} {\bibinfo {author} {\bibfnamefont {D.}~\bibnamefont
  {Blas}} \emph {et~al.},\ }\bibfield  {title} {\bibinfo {title} {{Large scale
  structure from viscous dark matter}},\ }\href
  {https://doi.org/10.1088/1475-7516/2015/11/049} {\bibfield  {journal}
  {\bibinfo  {journal} {JCAP}\ }\textbf {\bibinfo {volume} {11}},\ \bibinfo
  {pages} {049}},\ \Eprint {https://arxiv.org/abs/1507.06665} {arXiv:1507.06665
  [astro-ph.CO]} \BibitemShut {NoStop}%
\bibitem [{\citenamefont {Baumann}\ \emph {et~al.}(2012)\citenamefont
  {Baumann}, \citenamefont {Nicolis}, \citenamefont {Senatore},\ and\
  \citenamefont {Zaldarriaga}}]{Baumann:2010tm}%
  \BibitemOpen
  \bibfield  {author} {\bibinfo {author} {\bibfnamefont {D.}~\bibnamefont
  {Baumann}}, \bibinfo {author} {\bibfnamefont {A.}~\bibnamefont {Nicolis}},
  \bibinfo {author} {\bibfnamefont {L.}~\bibnamefont {Senatore}},\ and\
  \bibinfo {author} {\bibfnamefont {M.}~\bibnamefont {Zaldarriaga}},\
  }\bibfield  {title} {\bibinfo {title} {{Cosmological Non-Linearities as an
  Effective Fluid}},\ }\href {https://doi.org/10.1088/1475-7516/2012/07/051}
  {\bibfield  {journal} {\bibinfo  {journal} {JCAP}\ }\textbf {\bibinfo
  {volume} {07}},\ \bibinfo {pages} {051}},\ \Eprint
  {https://arxiv.org/abs/1004.2488} {arXiv:1004.2488 [astro-ph.CO]}
  \BibitemShut {NoStop}%
\bibitem [{\citenamefont {Carrasco}\ \emph {et~al.}(2012)\citenamefont
  {Carrasco}, \citenamefont {Hertzberg},\ and\ \citenamefont
  {Senatore}}]{Carrasco:2012cv}%
  \BibitemOpen
  \bibfield  {author} {\bibinfo {author} {\bibfnamefont {J.~J.~M.}\
  \bibnamefont {Carrasco}}, \bibinfo {author} {\bibfnamefont {M.~P.}\
  \bibnamefont {Hertzberg}},\ and\ \bibinfo {author} {\bibfnamefont
  {L.}~\bibnamefont {Senatore}},\ }\bibfield  {title} {\bibinfo {title} {{The
  Effective Field Theory of Cosmological Large Scale Structures}},\ }\href
  {https://doi.org/10.1007/JHEP09(2012)082} {\bibfield  {journal} {\bibinfo
  {journal} {JHEP}\ }\textbf {\bibinfo {volume} {09}},\ \bibinfo {pages}
  {082}},\ \Eprint {https://arxiv.org/abs/1206.2926} {arXiv:1206.2926
  [astro-ph.CO]} \BibitemShut {NoStop}%
\bibitem [{\citenamefont {Aric\`o}\ \emph {et~al.}(2021)\citenamefont
  {Aric\`o}, \citenamefont {Angulo},\ and\ \citenamefont
  {Zennaro}}]{Arico:2021izc}%
  \BibitemOpen
  \bibfield  {author} {\bibinfo {author} {\bibfnamefont {G.}~\bibnamefont
  {Aric\`o}}, \bibinfo {author} {\bibfnamefont {R.~E.}\ \bibnamefont
  {Angulo}},\ and\ \bibinfo {author} {\bibfnamefont {M.}~\bibnamefont
  {Zennaro}},\ }\bibfield  {title} {\bibinfo {title} {{Accelerating
  Large-Scale-Structure data analyses by emulating Boltzmann solvers and
  Lagrangian Perturbation Theory}}\ }\href
  {https://doi.org/10.12688/openreseurope.14310.2}
  {10.12688/openreseurope.14310.2} (\bibinfo {year} {2021}),\ \Eprint
  {https://arxiv.org/abs/2104.14568} {arXiv:2104.14568 [astro-ph.CO]}
  \BibitemShut {NoStop}%
\bibitem [{\citenamefont {Cusin}\ \emph {et~al.}(2018)\citenamefont {Cusin},
  \citenamefont {Lewandowski},\ and\ \citenamefont {Vernizzi}}]{Cusin:2017wjg}%
  \BibitemOpen
  \bibfield  {author} {\bibinfo {author} {\bibfnamefont {G.}~\bibnamefont
  {Cusin}}, \bibinfo {author} {\bibfnamefont {M.}~\bibnamefont {Lewandowski}},\
  and\ \bibinfo {author} {\bibfnamefont {F.}~\bibnamefont {Vernizzi}},\
  }\bibfield  {title} {\bibinfo {title} {{Dark Energy and Modified Gravity in
  the Effective Field Theory of Large-Scale Structure}},\ }\href
  {https://doi.org/10.1088/1475-7516/2018/04/005} {\bibfield  {journal}
  {\bibinfo  {journal} {JCAP}\ }\textbf {\bibinfo {volume} {04}},\ \bibinfo
  {pages} {005}},\ \Eprint {https://arxiv.org/abs/1712.02783} {arXiv:1712.02783
  [astro-ph.CO]} \BibitemShut {NoStop}%
\bibitem [{\citenamefont {Assassi}\ \emph {et~al.}(2015)\citenamefont {Assassi}
  \emph {et~al.}}]{Assassi:2015jqa}%
  \BibitemOpen
  \bibfield  {author} {\bibinfo {author} {\bibfnamefont {V.}~\bibnamefont
  {Assassi}} \emph {et~al.},\ }\bibfield  {title} {\bibinfo {title} {{Effective
  theory of large-scale structure with primordial non-Gaussianity}},\ }\href
  {https://doi.org/10.1088/1475-7516/2015/11/024} {\bibfield  {journal}
  {\bibinfo  {journal} {JCAP}\ }\textbf {\bibinfo {volume} {11}},\ \bibinfo
  {pages} {024}},\ \Eprint {https://arxiv.org/abs/1505.06668} {arXiv:1505.06668
  [astro-ph.CO]} \BibitemShut {NoStop}%
\bibitem [{\citenamefont {Senatore}\ and\ \citenamefont
  {Zaldarriaga}(2017)}]{Senatore:2017hyk}%
  \BibitemOpen
  \bibfield  {author} {\bibinfo {author} {\bibfnamefont {L.}~\bibnamefont
  {Senatore}}\ and\ \bibinfo {author} {\bibfnamefont {M.}~\bibnamefont
  {Zaldarriaga}},\ }\bibfield  {title} {\bibinfo {title} {{The Effective Field
  Theory of Large-Scale Structure in the presence of Massive Neutrinos}},\
  }\href@noop {} {\  (\bibinfo {year} {2017})},\ \Eprint
  {https://arxiv.org/abs/1707.04698} {arXiv:1707.04698 [astro-ph.CO]}
  \BibitemShut {NoStop}%
\bibitem [{\citenamefont {Chudaykin}\ and\ \citenamefont
  {Ivanov}(2019)}]{Chudaykin:2019ock}%
  \BibitemOpen
  \bibfield  {author} {\bibinfo {author} {\bibfnamefont {A.}~\bibnamefont
  {Chudaykin}}\ and\ \bibinfo {author} {\bibfnamefont {M.~M.}\ \bibnamefont
  {Ivanov}},\ }\bibfield  {title} {\bibinfo {title} {{Measuring neutrino masses
  with large-scale structure: Euclid forecast with controlled theoretical
  error}},\ }\href {https://doi.org/10.1088/1475-7516/2019/11/034} {\bibfield
  {journal} {\bibinfo  {journal} {JCAP}\ }\textbf {\bibinfo {volume} {11}},\
  \bibinfo {pages} {034}},\ \Eprint {https://arxiv.org/abs/1907.06666}
  {arXiv:1907.06666 [astro-ph.CO]} \BibitemShut {NoStop}%
\bibitem [{\citenamefont {Vasudevan}\ \emph {et~al.}(2019)\citenamefont
  {Vasudevan}, \citenamefont {Ivanov}, \citenamefont {Sibiryakov},\ and\
  \citenamefont {Lesgourgues}}]{Vasudevan:2019ewf}%
  \BibitemOpen
  \bibfield  {author} {\bibinfo {author} {\bibfnamefont {A.}~\bibnamefont
  {Vasudevan}}, \bibinfo {author} {\bibfnamefont {M.~M.}\ \bibnamefont
  {Ivanov}}, \bibinfo {author} {\bibfnamefont {S.}~\bibnamefont {Sibiryakov}},\
  and\ \bibinfo {author} {\bibfnamefont {J.}~\bibnamefont {Lesgourgues}},\
  }\bibfield  {title} {\bibinfo {title} {{Time-sliced perturbation theory with
  primordial non-Gaussianity and effects of large bulk flows on inflationary
  oscillating features}},\ }\href
  {https://doi.org/10.1088/1475-7516/2019/09/037} {\bibfield  {journal}
  {\bibinfo  {journal} {JCAP}\ }\textbf {\bibinfo {volume} {09}},\ \bibinfo
  {pages} {037}},\ \Eprint {https://arxiv.org/abs/1906.08697} {arXiv:1906.08697
  [astro-ph.CO]} \BibitemShut {NoStop}%
\bibitem [{\citenamefont {Zhang}\ and\ \citenamefont
  {Cai}(2022)}]{Zhang:2021uyp}%
  \BibitemOpen
  \bibfield  {author} {\bibinfo {author} {\bibfnamefont {P.}~\bibnamefont
  {Zhang}}\ and\ \bibinfo {author} {\bibfnamefont {Y.}~\bibnamefont {Cai}},\
  }\bibfield  {title} {\bibinfo {title} {{BOSS full-shape analysis from the
  EFTofLSS with exact time dependence}},\ }\href
  {https://doi.org/10.1088/1475-7516/2022/01/031} {\bibfield  {journal}
  {\bibinfo  {journal} {JCAP}\ }\textbf {\bibinfo {volume} {01}}\bibfield
  {number} {\bibinfo  {number} { (01)},\ \bibinfo {pages} {031}},\ }\Eprint
  {https://arxiv.org/abs/2111.05739} {arXiv:2111.05739 [astro-ph.CO]}
  \BibitemShut {NoStop}%
\bibitem [{\citenamefont {Ivanov}\ \emph {et~al.}(2020)\citenamefont {Ivanov},
  \citenamefont {Simonovi\'c},\ and\ \citenamefont
  {Zaldarriaga}}]{Ivanov:2019pdj}%
  \BibitemOpen
  \bibfield  {author} {\bibinfo {author} {\bibfnamefont {M.~M.}\ \bibnamefont
  {Ivanov}}, \bibinfo {author} {\bibfnamefont {M.}~\bibnamefont
  {Simonovi\'c}},\ and\ \bibinfo {author} {\bibfnamefont {M.}~\bibnamefont
  {Zaldarriaga}},\ }\bibfield  {title} {\bibinfo {title} {{Cosmological
  Parameters from the BOSS Galaxy Power Spectrum}},\ }\href
  {https://doi.org/10.1088/1475-7516/2020/05/042} {\bibfield  {journal}
  {\bibinfo  {journal} {JCAP}\ }\textbf {\bibinfo {volume} {05}},\ \bibinfo
  {pages} {042}},\ \Eprint {https://arxiv.org/abs/1909.05277} {arXiv:1909.05277
  [astro-ph.CO]} \BibitemShut {NoStop}%
\bibitem [{\citenamefont {Carrilho}\ \emph {et~al.}(2023)\citenamefont
  {Carrilho}, \citenamefont {Moretti},\ and\ \citenamefont
  {Pourtsidou}}]{Carrilho:2022mon}%
  \BibitemOpen
  \bibfield  {author} {\bibinfo {author} {\bibfnamefont {P.}~\bibnamefont
  {Carrilho}}, \bibinfo {author} {\bibfnamefont {C.}~\bibnamefont {Moretti}},\
  and\ \bibinfo {author} {\bibfnamefont {A.}~\bibnamefont {Pourtsidou}},\
  }\bibfield  {title} {\bibinfo {title} {{Cosmology with the EFTofLSS and BOSS:
  dark energy constraints and a note on priors}},\ }\href
  {https://doi.org/10.1088/1475-7516/2023/01/028} {\bibfield  {journal}
  {\bibinfo  {journal} {JCAP}\ }\textbf {\bibinfo {volume} {01}},\ \bibinfo
  {pages} {028}},\ \Eprint {https://arxiv.org/abs/2207.14784} {arXiv:2207.14784
  [astro-ph.CO]} \BibitemShut {NoStop}%
\bibitem [{\citenamefont {Adame}\ \emph {et~al.}(2024)\citenamefont {Adame}
  \emph {et~al.}}]{DESI:2024mwx}%
  \BibitemOpen
  \bibfield  {author} {\bibinfo {author} {\bibfnamefont {A.~G.}\ \bibnamefont
  {Adame}} \emph {et~al.} (\bibinfo {collaboration} {DESI}),\ }\bibfield
  {title} {\bibinfo {title} {{DESI 2024 VI: Cosmological Constraints from the
  Measurements of Baryon Acoustic Oscillations}},\ }\href@noop {} {\  (\bibinfo
  {year} {2024})},\ \Eprint {https://arxiv.org/abs/2404.03002}
  {arXiv:2404.03002 [astro-ph.CO]} \BibitemShut {NoStop}%
\bibitem [{\citenamefont {Chen}\ \emph {et~al.}(2024)\citenamefont {Chen} \emph
  {et~al.}}]{Chen:2024vuf}%
  \BibitemOpen
  \bibfield  {author} {\bibinfo {author} {\bibfnamefont {S.-F.}\ \bibnamefont
  {Chen}} \emph {et~al.},\ }\bibfield  {title} {\bibinfo {title} {{Suppression
  without Thawing: Constraining Structure Formation and Dark Energy with Galaxy
  Clustering}},\ }\href@noop {} {\  (\bibinfo {year} {2024})},\ \Eprint
  {https://arxiv.org/abs/2406.13388} {arXiv:2406.13388 [astro-ph.CO]}
  \BibitemShut {NoStop}%
\bibitem [{\citenamefont {Cabass}\ \emph {et~al.}(2024)\citenamefont {Cabass}
  \emph {et~al.}}]{Cabass:2024wob}%
  \BibitemOpen
  \bibfield  {author} {\bibinfo {author} {\bibfnamefont {G.}~\bibnamefont
  {Cabass}} \emph {et~al.},\ }\bibfield  {title} {\bibinfo {title} {{BOSS
  Constraints on Massive Particles during Inflation: The Cosmological Collider
  in Action}},\ }\href@noop {} {\  (\bibinfo {year} {2024})},\ \Eprint
  {https://arxiv.org/abs/2404.01894} {arXiv:2404.01894 [astro-ph.CO]}
  \BibitemShut {NoStop}%
\bibitem [{\citenamefont {Sugiyama}\ \emph {et~al.}(2023)\citenamefont
  {Sugiyama} \emph {et~al.}}]{Sugiyama:2023tes}%
  \BibitemOpen
  \bibfield  {author} {\bibinfo {author} {\bibfnamefont {N.~S.}\ \bibnamefont
  {Sugiyama}} \emph {et~al.},\ }\bibfield  {title} {\bibinfo {title} {{New
  constraints on cosmological modified gravity theories from anisotropic
  three-point correlation functions of BOSS DR12 galaxies}}\ }\href
  {https://doi.org/10.1093/mnras/stad1505} {10.1093/mnras/stad1505} (\bibinfo
  {year} {2023}),\ \Eprint {https://arxiv.org/abs/2302.06808} {arXiv:2302.06808
  [astro-ph.CO]} \BibitemShut {NoStop}%
\bibitem [{\citenamefont {Hoyle}\ \emph {et~al.}(2018)\citenamefont {Hoyle}
  \emph {et~al.}}]{DES:2017ndt}%
  \BibitemOpen
  \bibfield  {author} {\bibinfo {author} {\bibfnamefont {B.}~\bibnamefont
  {Hoyle}} \emph {et~al.} (\bibinfo {collaboration} {DES}),\ }\bibfield
  {title} {\bibinfo {title} {{Dark Energy Survey Year 1 Results: Redshift
  distributions of the weak lensing source galaxies}},\ }\href
  {https://doi.org/10.1093/mnras/sty957} {\bibfield  {journal} {\bibinfo
  {journal} {Mon. Not. Roy. Astron. Soc.}\ }\textbf {\bibinfo {volume} {478}},\
  \bibinfo {pages} {592} (\bibinfo {year} {2018})},\ \Eprint
  {https://arxiv.org/abs/1708.01532} {arXiv:1708.01532 [astro-ph.CO]}
  \BibitemShut {NoStop}%
\bibitem [{\citenamefont {Ilbert}\ \emph {et~al.}(2021)\citenamefont {Ilbert}
  \emph {et~al.}}]{Euclid:2021upd}%
  \BibitemOpen
  \bibfield  {author} {\bibinfo {author} {\bibfnamefont {O.}~\bibnamefont
  {Ilbert}} \emph {et~al.} (\bibinfo {collaboration} {Euclid}),\ }\bibfield
  {title} {\bibinfo {title} {{Euclid preparation. XI. Mean redshift
  determination from galaxy redshift probabilities for cosmic shear
  tomography}},\ }\href {https://doi.org/10.1051/0004-6361/202040237}
  {\bibfield  {journal} {\bibinfo  {journal} {Astron. Astrophys.}\ }\textbf
  {\bibinfo {volume} {647}},\ \bibinfo {pages} {A117} (\bibinfo {year}
  {2021})},\ \Eprint {https://arxiv.org/abs/2101.02228} {arXiv:2101.02228
  [astro-ph.CO]} \BibitemShut {NoStop}%
\bibitem [{\citenamefont {{Bolton}}\ \emph {et~al.}(2012)\citenamefont
  {{Bolton}} \emph {et~al.}}]{2012AJ....144..144B}%
  \BibitemOpen
  \bibfield  {author} {\bibinfo {author} {\bibfnamefont {A.~S.}\ \bibnamefont
  {{Bolton}}} \emph {et~al.},\ }\bibfield  {title} {\bibinfo {title} {{Spectral
  Classification and Redshift Measurement for the SDSS-III Baryon Oscillation
  Spectroscopic Survey}},\ }\href {https://doi.org/10.1088/0004-6256/144/5/144}
  {\bibfield  {journal} {\bibinfo  {journal} {The Astronomical Journal}\
  }\textbf {\bibinfo {volume} {144}},\ \bibinfo {eid} {144} (\bibinfo {year}
  {2012})},\ \Eprint {https://arxiv.org/abs/1207.7326} {arXiv:1207.7326
  [astro-ph.CO]} \BibitemShut {NoStop}%
\bibitem [{\citenamefont {Scoccimarro}\ \emph {et~al.}(1998)\citenamefont
  {Scoccimarro}, \citenamefont {Colombi}, \citenamefont {Fry}, \citenamefont
  {Frieman}, \citenamefont {Hivon},\ and\ \citenamefont
  {Melott}}]{Scoccimarro:1997st}%
  \BibitemOpen
  \bibfield  {author} {\bibinfo {author} {\bibfnamefont {R.}~\bibnamefont
  {Scoccimarro}}, \bibinfo {author} {\bibfnamefont {S.}~\bibnamefont
  {Colombi}}, \bibinfo {author} {\bibfnamefont {J.~N.}\ \bibnamefont {Fry}},
  \bibinfo {author} {\bibfnamefont {J.~A.}\ \bibnamefont {Frieman}}, \bibinfo
  {author} {\bibfnamefont {E.}~\bibnamefont {Hivon}},\ and\ \bibinfo {author}
  {\bibfnamefont {A.}~\bibnamefont {Melott}},\ }\bibfield  {title} {\bibinfo
  {title} {{Nonlinear evolution of the bispectrum of cosmological
  perturbations}},\ }\href {https://doi.org/10.1086/305399} {\bibfield
  {journal} {\bibinfo  {journal} {Astrophys. J.}\ }\textbf {\bibinfo {volume}
  {496}},\ \bibinfo {pages} {586} (\bibinfo {year} {1998})},\ \Eprint
  {https://arxiv.org/abs/astro-ph/9704075} {arXiv:astro-ph/9704075}
  \BibitemShut {NoStop}%
\bibitem [{\citenamefont {Pryer}\ \emph {et~al.}(2022)\citenamefont {Pryer},
  \citenamefont {Smith}, \citenamefont {Booth}, \citenamefont {Blake},
  \citenamefont {Eggemeier},\ and\ \citenamefont {Loveday}}]{Pryer:2021cut}%
  \BibitemOpen
  \bibfield  {author} {\bibinfo {author} {\bibfnamefont {D.}~\bibnamefont
  {Pryer}}, \bibinfo {author} {\bibfnamefont {R.~E.}\ \bibnamefont {Smith}},
  \bibinfo {author} {\bibfnamefont {R.}~\bibnamefont {Booth}}, \bibinfo
  {author} {\bibfnamefont {C.}~\bibnamefont {Blake}}, \bibinfo {author}
  {\bibfnamefont {A.}~\bibnamefont {Eggemeier}},\ and\ \bibinfo {author}
  {\bibfnamefont {J.}~\bibnamefont {Loveday}},\ }\bibfield  {title} {\bibinfo
  {title} {{The galaxy power spectrum on the lightcone: deep, wide-angle
  redshift surveys and the turnover scale}},\ }\href
  {https://doi.org/10.1088/1475-7516/2022/08/019} {\bibfield  {journal}
  {\bibinfo  {journal} {JCAP}\ }\textbf {\bibinfo {volume} {08}}\bibfield
  {number} {\bibinfo  {number} { (08)},\ \bibinfo {pages} {019}},\ }\Eprint
  {https://arxiv.org/abs/2111.01811} {arXiv:2111.01811 [astro-ph.CO]}
  \BibitemShut {NoStop}%
\bibitem [{\citenamefont {Semenzato}\ \emph {et~al.}(2024)\citenamefont
  {Semenzato}, \citenamefont {Bertacca},\ and\ \citenamefont
  {Raccanelli}}]{Semenzato:2024rlc}%
  \BibitemOpen
  \bibfield  {author} {\bibinfo {author} {\bibfnamefont {F.}~\bibnamefont
  {Semenzato}}, \bibinfo {author} {\bibfnamefont {D.}~\bibnamefont
  {Bertacca}},\ and\ \bibinfo {author} {\bibfnamefont {A.}~\bibnamefont
  {Raccanelli}},\ }\bibfield  {title} {\bibinfo {title} {{The full-sky
  Spherical Fourier-Bessel power spectrum in general relativity}},\ }\href@noop
  {} {\  (\bibinfo {year} {2024})},\ \Eprint {https://arxiv.org/abs/2406.09545}
  {arXiv:2406.09545 [astro-ph.CO]} \BibitemShut {NoStop}%
\bibitem [{\citenamefont {Raccanelli}\ and\ \citenamefont
  {Vlah}(2023{\natexlab{a}})}]{Raccanelli:2023fle}%
  \BibitemOpen
  \bibfield  {author} {\bibinfo {author} {\bibfnamefont {A.}~\bibnamefont
  {Raccanelli}}\ and\ \bibinfo {author} {\bibfnamefont {Z.}~\bibnamefont
  {Vlah}},\ }\bibfield  {title} {\bibinfo {title} {{Power spectrum in the
  cave}},\ }\href@noop {} {\  (\bibinfo {year} {2023}{\natexlab{a}})},\ \Eprint
  {https://arxiv.org/abs/2305.16278} {arXiv:2305.16278 [astro-ph.CO]}
  \BibitemShut {NoStop}%
\bibitem [{\citenamefont {Raccanelli}\ and\ \citenamefont
  {Vlah}(2023{\natexlab{b}})}]{Raccanelli:2023zkj}%
  \BibitemOpen
  \bibfield  {author} {\bibinfo {author} {\bibfnamefont {A.}~\bibnamefont
  {Raccanelli}}\ and\ \bibinfo {author} {\bibfnamefont {Z.}~\bibnamefont
  {Vlah}},\ }\bibfield  {title} {\bibinfo {title} {{Observed power spectrum and
  frequency-angular power spectrum}},\ }\href
  {https://doi.org/10.1103/PhysRevD.108.043537} {\bibfield  {journal} {\bibinfo
   {journal} {Phys. Rev. D}\ }\textbf {\bibinfo {volume} {108}},\ \bibinfo
  {pages} {043537} (\bibinfo {year} {2023}{\natexlab{b}})},\ \Eprint
  {https://arxiv.org/abs/2306.00808} {arXiv:2306.00808 [astro-ph.CO]}
  \BibitemShut {NoStop}%
\bibitem [{\citenamefont {Gao}\ \emph {et~al.}(2023)\citenamefont {Gao},
  \citenamefont {Raccanelli},\ and\ \citenamefont {Vlah}}]{Gao:2023rmo}%
  \BibitemOpen
  \bibfield  {author} {\bibinfo {author} {\bibfnamefont {Z.}~\bibnamefont
  {Gao}}, \bibinfo {author} {\bibfnamefont {A.}~\bibnamefont {Raccanelli}},\
  and\ \bibinfo {author} {\bibfnamefont {Z.}~\bibnamefont {Vlah}},\ }\bibfield
  {title} {\bibinfo {title} {{Asymptotic connection between full- and flat-sky
  angular correlators}},\ }\href {https://doi.org/10.1103/PhysRevD.108.043503}
  {\bibfield  {journal} {\bibinfo  {journal} {Phys. Rev. D}\ }\textbf {\bibinfo
  {volume} {108}},\ \bibinfo {pages} {043503} (\bibinfo {year} {2023})},\
  \Eprint {https://arxiv.org/abs/2306.02993} {arXiv:2306.02993 [astro-ph.CO]}
  \BibitemShut {NoStop}%
\bibitem [{\citenamefont {Gao}\ \emph {et~al.}(2024)\citenamefont {Gao},
  \citenamefont {Vlah},\ and\ \citenamefont {Challinor}}]{Gao:2023tcd}%
  \BibitemOpen
  \bibfield  {author} {\bibinfo {author} {\bibfnamefont {Z.}~\bibnamefont
  {Gao}}, \bibinfo {author} {\bibfnamefont {Z.}~\bibnamefont {Vlah}},\ and\
  \bibinfo {author} {\bibfnamefont {A.}~\bibnamefont {Challinor}},\ }\bibfield
  {title} {\bibinfo {title} {{Flat-sky angular power spectra revisited}},\
  }\href {https://doi.org/10.1088/1475-7516/2024/02/003} {\bibfield  {journal}
  {\bibinfo  {journal} {JCAP}\ }\textbf {\bibinfo {volume} {02}},\ \bibinfo
  {pages} {003}},\ \Eprint {https://arxiv.org/abs/2307.13768} {arXiv:2307.13768
  [astro-ph.CO]} \BibitemShut {NoStop}%
\bibitem [{\citenamefont {Desjacques}\ \emph {et~al.}(2018)\citenamefont
  {Desjacques}, \citenamefont {Jeong},\ and\ \citenamefont
  {Schmidt}}]{Desjacques:2016bnm}%
  \BibitemOpen
  \bibfield  {author} {\bibinfo {author} {\bibfnamefont {V.}~\bibnamefont
  {Desjacques}}, \bibinfo {author} {\bibfnamefont {D.}~\bibnamefont {Jeong}},\
  and\ \bibinfo {author} {\bibfnamefont {F.}~\bibnamefont {Schmidt}},\
  }\bibfield  {title} {\bibinfo {title} {{Large-Scale Galaxy Bias}},\ }\href
  {https://doi.org/10.1016/j.physrep.2017.12.002} {\bibfield  {journal}
  {\bibinfo  {journal} {Phys. Rept.}\ }\textbf {\bibinfo {volume} {733}},\
  \bibinfo {pages} {1} (\bibinfo {year} {2018})},\ \Eprint
  {https://arxiv.org/abs/1611.09787} {arXiv:1611.09787 [astro-ph.CO]}
  \BibitemShut {NoStop}%
\bibitem [{\citenamefont {Tucker}\ \emph {et~al.}(1997)\citenamefont {Tucker}
  \emph {et~al.}}]{10.1093/mnras/285.1.L5}%
  \BibitemOpen
  \bibfield  {author} {\bibinfo {author} {\bibfnamefont {D.~L.}\ \bibnamefont
  {Tucker}} \emph {et~al.},\ }\bibfield  {title} {\bibinfo {title} {{The Las
  Campanas Redshift Survey galaxy—galaxy autocorrelation function}},\ }\href
  {https://doi.org/10.1093/mnras/285.1.L5} {\bibfield  {journal} {\bibinfo
  {journal} {Monthly Notices of the Royal Astronomical Society}\ }\textbf
  {\bibinfo {volume} {285}},\ \bibinfo {pages} {L5} (\bibinfo {year} {1997})},\
  \Eprint
  {https://arxiv.org/abs/https://academic.oup.com/mnras/article-pdf/285/1/L5/3158615/285-1-L5.pdf}
  {https://academic.oup.com/mnras/article-pdf/285/1/L5/3158615/285-1-L5.pdf}
  \BibitemShut {NoStop}%
\bibitem [{\citenamefont {{Krause}}\ and\ \citenamefont
  {{Hirata}}()}]{2010A&A...523A..28K}%
  \BibitemOpen
  \bibfield  {author} {\bibinfo {author} {\bibfnamefont {E.}~\bibnamefont
  {{Krause}}}\ and\ \bibinfo {author} {\bibfnamefont {C.~M.}\ \bibnamefont
  {{Hirata}}},\ }\bibfield  {title} {\bibinfo {title} {{Weak lensing power
  spectra for precision cosmology. Multiple-deflection, reduced shear, and
  lensing bias corrections}},\ }\href@noop {} {\bibinfo  {journal} {Astronomy
  and Astrophysics}\ }\BibitemShut {NoStop}%
\bibitem [{\citenamefont {{Deshpande}}\ \emph {et~al.}(2024)\citenamefont
  {{Deshpande}} \emph {et~al.}}]{2024A&A...684A.138E}%
  \BibitemOpen
\bibfield  {journal} {  }\bibfield  {author} {\bibinfo {author} {\bibfnamefont
  {A.~C.}\ \bibnamefont {{Deshpande}}} \emph {et~al.},\ }\bibfield  {title}
  {\bibinfo {title} {{Euclid preparation. XXXVI. Modelling the weak lensing
  angular power spectrum}},\ }\href
  {https://doi.org/10.1051/0004-6361/202346110} {\bibfield  {journal} {\bibinfo
   {journal} {Astronomy and Astrophysics}\ }\textbf {\bibinfo {volume} {684}},\
  \bibinfo {eid} {A138} (\bibinfo {year} {2024})},\ \Eprint
  {https://arxiv.org/abs/2302.04507} {arXiv:2302.04507 [astro-ph.CO]}
  \BibitemShut {NoStop}%
\bibitem [{\citenamefont {Hayashi}\ and\ \citenamefont
  {White}(2008)}]{10.1111/j.1365-2966.2008.13371.x}%
  \BibitemOpen
  \bibfield  {author} {\bibinfo {author} {\bibfnamefont {E.}~\bibnamefont
  {Hayashi}}\ and\ \bibinfo {author} {\bibfnamefont {S.~D.~M.}\ \bibnamefont
  {White}},\ }\bibfield  {title} {\bibinfo {title} {{Understanding the
  halo-mass and galaxy-mass cross-correlation functions}},\ }\href
  {https://doi.org/10.1111/j.1365-2966.2008.13371.x} {\bibfield  {journal}
  {\bibinfo  {journal} {Monthly Notices of the Royal Astronomical Society}\
  }\textbf {\bibinfo {volume} {388}},\ \bibinfo {pages} {2} (\bibinfo {year}
  {2008})},\ \Eprint
  {https://arxiv.org/abs/https://academic.oup.com/mnras/article-pdf/388/1/2/18718120/mnras0388-0002.pdf}
  {https://academic.oup.com/mnras/article-pdf/388/1/2/18718120/mnras0388-0002.pdf}
  \BibitemShut {NoStop}%
\bibitem [{\citenamefont {Fraser}\ \emph {et~al.}(2024)\citenamefont {Fraser},
  \citenamefont {Paillas}, \citenamefont {Percival}, \citenamefont {Nadathur},
  \citenamefont {Radinovi\'c},\ and\ \citenamefont {Winther}}]{Fraser:2024ecp}%
  \BibitemOpen
  \bibfield  {author} {\bibinfo {author} {\bibfnamefont {T.~S.}\ \bibnamefont
  {Fraser}}, \bibinfo {author} {\bibfnamefont {E.}~\bibnamefont {Paillas}},
  \bibinfo {author} {\bibfnamefont {W.~J.}\ \bibnamefont {Percival}}, \bibinfo
  {author} {\bibfnamefont {S.}~\bibnamefont {Nadathur}}, \bibinfo {author}
  {\bibfnamefont {S.}~\bibnamefont {Radinovi\'c}},\ and\ \bibinfo {author}
  {\bibfnamefont {H.~A.}\ \bibnamefont {Winther}},\ }\bibfield  {title}
  {\bibinfo {title} {{Modelling the BOSS void-galaxy cross-correlation function
  using a neural-network emulator}},\ }\href@noop {} {\  (\bibinfo {year}
  {2024})},\ \Eprint {https://arxiv.org/abs/2407.03221} {arXiv:2407.03221
  [astro-ph.CO]} \BibitemShut {NoStop}%
\bibitem [{\citenamefont {Kirk}\ \emph {et~al.}(2015)\citenamefont {Kirk} \emph
  {et~al.}}]{Kirk:2015xqa}%
  \BibitemOpen
  \bibfield  {author} {\bibinfo {author} {\bibfnamefont {D.}~\bibnamefont
  {Kirk}} \emph {et~al.},\ }\bibfield  {title} {\bibinfo {title} {{Cross
  correlation surveys with the Square Kilometre Array}},\ }\href@noop {} {\
  (\bibinfo {year} {2015})},\ \Eprint {https://arxiv.org/abs/1501.03848}
  {arXiv:1501.03848 [astro-ph.CO]} \BibitemShut {NoStop}%
\bibitem [{\citenamefont {Dam}\ \emph {et~al.}(2021)\citenamefont {Dam},
  \citenamefont {Bolejko},\ and\ \citenamefont {Lewis}}]{Dam:2021fff}%
  \BibitemOpen
  \bibfield  {author} {\bibinfo {author} {\bibfnamefont {L.}~\bibnamefont
  {Dam}}, \bibinfo {author} {\bibfnamefont {K.}~\bibnamefont {Bolejko}},\ and\
  \bibinfo {author} {\bibfnamefont {G.~F.}\ \bibnamefont {Lewis}},\ }\bibfield
  {title} {\bibinfo {title} {{Exploring the redshift-space peculiar velocity
  field and its power spectrum}},\ }\href
  {https://doi.org/10.1088/1475-7516/2021/09/018} {\bibfield  {journal}
  {\bibinfo  {journal} {JCAP}\ }\textbf {\bibinfo {volume} {09}},\ \bibinfo
  {pages} {018}},\ \Eprint {https://arxiv.org/abs/2105.12933} {arXiv:2105.12933
  [astro-ph.CO]} \BibitemShut {NoStop}%
\bibitem [{\citenamefont {Howlett}\ \emph {et~al.}(2017)\citenamefont {Howlett}
  \emph {et~al.}}]{Howlett:2017asq}%
  \BibitemOpen
  \bibfield  {author} {\bibinfo {author} {\bibfnamefont {C.}~\bibnamefont
  {Howlett}} \emph {et~al.},\ }\bibfield  {title} {\bibinfo {title} {{2MTF
  \textendash{} VI. Measuring the velocity power spectrum}},\ }\href
  {https://doi.org/10.1093/mnras/stx1521} {\bibfield  {journal} {\bibinfo
  {journal} {Mon. Not. Roy. Astron. Soc.}\ }\textbf {\bibinfo {volume} {471}},\
  \bibinfo {pages} {3135} (\bibinfo {year} {2017})},\ \Eprint
  {https://arxiv.org/abs/1706.05130} {arXiv:1706.05130 [astro-ph.CO]}
  \BibitemShut {NoStop}%
\bibitem [{\citenamefont {Tonegawa}\ \emph {et~al.}(2024)\citenamefont
  {Tonegawa} \emph {et~al.}}]{Tonegawa:2023gbf}%
  \BibitemOpen
  \bibfield  {author} {\bibinfo {author} {\bibfnamefont {M.}~\bibnamefont
  {Tonegawa}} \emph {et~al.},\ }\bibfield  {title} {\bibinfo {title} {{The
  effects of non-linearity on the growth rate constraint from velocity
  correlation functions}},\ }\href {https://doi.org/10.1093/mnras/stae700}
  {\bibfield  {journal} {\bibinfo  {journal} {Mon. Not. Roy. Astron. Soc.}\
  }\textbf {\bibinfo {volume} {529}},\ \bibinfo {pages} {4787} (\bibinfo {year}
  {2024})},\ \Eprint {https://arxiv.org/abs/2309.14457} {arXiv:2309.14457
  [astro-ph.CO]} \BibitemShut {NoStop}%
\bibitem [{\citenamefont {Kacprzak}\ and\ \citenamefont
  {Fluri}(2022)}]{Kacprzak:2022oit}%
  \BibitemOpen
  \bibfield  {author} {\bibinfo {author} {\bibfnamefont {T.}~\bibnamefont
  {Kacprzak}}\ and\ \bibinfo {author} {\bibfnamefont {J.}~\bibnamefont
  {Fluri}},\ }\bibfield  {title} {\bibinfo {title} {{DeepLSS: Breaking
  Parameter Degeneracies in Large-Scale Structure with Deep-Learning Analysis
  of Combined Probes}},\ }\href {https://doi.org/10.1103/PhysRevX.12.031029}
  {\bibfield  {journal} {\bibinfo  {journal} {Phys. Rev. X}\ }\textbf {\bibinfo
  {volume} {12}},\ \bibinfo {pages} {031029} (\bibinfo {year} {2022})},\
  \Eprint {https://arxiv.org/abs/2203.09616} {arXiv:2203.09616 [astro-ph.CO]}
  \BibitemShut {NoStop}%
\bibitem [{\citenamefont {Singh}\ \emph {et~al.}(2020)\citenamefont {Singh}
  \emph {et~al.}}]{Singh:2018kmr}%
  \BibitemOpen
  \bibfield  {author} {\bibinfo {author} {\bibfnamefont {S.}~\bibnamefont
  {Singh}} \emph {et~al.},\ }\bibfield  {title} {\bibinfo {title}
  {{Cosmological constraints from galaxy lensing cross-correlations using BOSS
  galaxies with SDSS and CMB lensing}},\ }\href
  {https://doi.org/10.1093/mnras/stz2922} {\bibfield  {journal} {\bibinfo
  {journal} {Mon. Not. Roy. Astron. Soc.}\ }\textbf {\bibinfo {volume} {491}},\
  \bibinfo {pages} {51} (\bibinfo {year} {2020})},\ \Eprint
  {https://arxiv.org/abs/1811.06499} {arXiv:1811.06499 [astro-ph.CO]}
  \BibitemShut {NoStop}%
\bibitem [{\citenamefont {Schmittfull}\ and\ \citenamefont
  {Seljak}(2018)}]{Schmittfull:2017ffw}%
  \BibitemOpen
  \bibfield  {author} {\bibinfo {author} {\bibfnamefont {M.}~\bibnamefont
  {Schmittfull}}\ and\ \bibinfo {author} {\bibfnamefont {U.}~\bibnamefont
  {Seljak}},\ }\bibfield  {title} {\bibinfo {title} {{Parameter constraints
  from cross-correlation of CMB lensing with galaxy clustering}},\ }\href
  {https://doi.org/10.1103/PhysRevD.97.123540} {\bibfield  {journal} {\bibinfo
  {journal} {Phys. Rev. D}\ }\textbf {\bibinfo {volume} {97}},\ \bibinfo
  {pages} {123540} (\bibinfo {year} {2018})},\ \Eprint
  {https://arxiv.org/abs/1710.09465} {arXiv:1710.09465 [astro-ph.CO]}
  \BibitemShut {NoStop}%
\bibitem [{\citenamefont {Scoccimarro}(2015)}]{Scoccimarro:2015bla}%
  \BibitemOpen
  \bibfield  {author} {\bibinfo {author} {\bibfnamefont {R.}~\bibnamefont
  {Scoccimarro}},\ }\bibfield  {title} {\bibinfo {title} {{Fast Estimators for
  Redshift-Space Clustering}},\ }\href
  {https://doi.org/10.1103/PhysRevD.92.083532} {\bibfield  {journal} {\bibinfo
  {journal} {Phys. Rev. D}\ }\textbf {\bibinfo {volume} {92}},\ \bibinfo
  {pages} {083532} (\bibinfo {year} {2015})},\ \Eprint
  {https://arxiv.org/abs/1506.02729} {arXiv:1506.02729 [astro-ph.CO]}
  \BibitemShut {NoStop}%
\end{thebibliography}%

\end{document}